\documentclass[11pt,a4paper]{article}
\usepackage{graphicx} 
\usepackage{xcolor}
\usepackage{a4wide}
\usepackage{amssymb}
\usepackage{amsmath}
\usepackage{comment}
\usepackage{algorithm}
\usepackage[noend]{algpseudocode}
\usepackage{stmaryrd}
\usepackage{xspace}
\usepackage{authblk}
\usepackage[normalem]{ulem}
\usepackage{multicol}
\usepackage{soul}
\usepackage{jheppub} 

\renewcommand{\d}{\ensuremath{\mathrm{d}}}
\newcommand{\Li}{\ensuremath{\mathrm{Li}_2}}

\newcommand{\mapPhi}{\ensuremath{\hat{\Phi}}}
\newcommand{\calB}{\ensuremath{{\mathsf{ B}}}}
\newcommand{\calT}{\ensuremath{{\mathsf{ T}}}}
\newcommand{\calV}{\ensuremath{{\mathsf{ V}}}}
\newcommand{\calW}{\ensuremath{{\mathsf{ W}}}}
\newcommand{\Ds}{\ensuremath{D_3^{{\rm s}0}}}
\newcommand{\qbar}{\ensuremath{\bar{q}}}
\newcommand{\mZ}{\ensuremath{m_Z}}
\newcommand{\KNLO}{\ensuremath{v^{\mathrm{NLO}}}}
\newcommand{\powheg}{POWHEG}

\newcounter{BEQ}

\newcommand{\eqsref}[1]{eqs.~\eqref{#1}\xspace}
\newcommand{\Secref}[1]{Sec.~\ref{#1}\xspace}
\newcommand{\secref}[1]{sec.~\ref{#1}\xspace}

\newcommand{\figref}[1]{fig.~\ref{#1}\xspace}

\newcommand{\bb}[1]{{\color{blue}#1}}

\newcommand{\yqg}{\hat{y}_{qg}}
\newcommand{\yqbg}{\hat{y}_{g\bar{q}}}

\usepackage{pifont}

\title{Matching Z $\to$ Hadrons at NNLO with Sector Showers}

\author[1]{Basem Kamal El-Menoufi,}
\author[2]{Christian T. Preuss,}
\author[1]{Ludovic Scyboz,}
\author[1]{Peter Skands}
\affiliation[1]{School of Physics \& Astronomy, Monash University, Clayton VIC-3800, Australia}
\affiliation[2]{University of Wuppertal, Department of Physics, DE-42119 Wuppertal, Germany}

\date{Received: date / Accepted: \today}

\abstract{
We present a detailed technical derivation of matching conditions at next-to-next-to-leading order in the sectorised VINCIA parton shower, by considering leading-colour 2-, 3- and 4-jet rates in hadronic Z-boson decays.
In particular, we introduce a full subtraction-based calculation of the matching coefficient required to obtain the NLO 3-jet rate. This is achieved through a judicious choice of the counter-terms, which optimises the numerical evaluation of the subtracted double-real matrix element. We additionally give a consistent prescription for incorporating interference effects due to higher-order mixing between Born states with different flavour contents.
Finally, we briefly comment on higher-order uncertainty estimates.

}

\keywords{QCD, Parton Shower, NNLO, Matching, LEP
  \\[4em]
  \textit{For the purpose of Open Access, the authors have applied a CC BY
  public copyright licence to any Author Accepted Manuscript (AAM)
  version arising from this submission.}
}

\begin{document}
\maketitle

\section{Introduction}

With ever-increasing precision on the side of collider experiments, it has become
imperative to address the accuracy of ingredients used in Monte Carlo event generators
(MCEG), in particular their parton-shower component. The latter is responsible for
generating the radiation pattern of QCD from the energetic scale of the hard collision
down to the hadronisation scale, and a precise description of this extra radiation is
crucial for the interpretation of data at the Large Hadron Collider (LHC).  Recently,
considerable efforts to improve the parton-shower component of MCEG’s have led to the
advent of next-to-leading logarithmic (NLL) accurate
showers~\cite{Dasgupta:2020fwr,vanBeekveld:2022ukn,vanBeekveld:2023chs,Forshaw:2020wrq,Nagy:2020dvz,Nagy:2020rmk,Herren:2022jej,Assi:2023rbu,Preuss:2024vyu},
with a clear path towards full NNLL
accuracy~\cite{FerrarioRavasio:2023kyg,vanBeekveld:2024wws}.

In addition to all-order resummation, the shower needs to be matched to fixed-order
calculations to correctly describe hard (resolved) radiation.  Two well-known schemes are
used to match parton showers at next-to-leading order (NLO):
MC@NLO~\cite{Frixione:2002ik} and \powheg~\cite{Nason:2004rx,Frixione:2007vw}, along with
a few more recent proposals~\cite{Jadach:2015mza,Sarmah:2024hdk,Nason:2021xke,vanBeekveld:2025lpz}.
Next-to-next-to-leading order (NNLO) matching is more involved, and several approaches
have been applied to specific processes of interest at the LHC,
i.e.~NNLOPS~\cite{Hamilton:2013fea}, UN$^2$LOPS~\cite{Lonnblad:2012ix,Hoche:2014uhw}, as
well as MiNNLO$_{\textrm{PS}}$~\cite{Monni:2019whf} and GENEVA~\cite{Alioli:2013hqa}.

Most existing schemes rely on the use of an external generator
to produce the first
hard emission(s), followed by a handover to the parton shower for the remaining emissions.
In order to retain the logarithmic accuracy of the shower, this type of approach requires a careful treatment in the handover, e.g.~alignment of contours
in the evolution variables of the hard generator and the shower (see refs.~\cite{Corke:2010zj,Hamilton:2023dwb}),
often in the form of vetoes~\cite{Catani:2001cc,Nason:2004rx}. Therefore it is more advantageous
to use the parton shower itself as a phase-space generator, and correct the shower radiation probability
for hard emission from fixed-order matrix elements~\cite{Hoche:2010pf,Platzer:2011bc}.
This approach crucially relies on the fact that the parton shower fully covers the available phase space.

Recently, a proposal for matching \textit{sectorised} parton showers to NNLO perturbative accuracy
was presented by several of us in ref.~\cite{Campbell:2021svd}.
Sector showers, as implemented e.g.~in VINCIA~\cite{Lopez-Villarejo:2011pwr,Larkoski:2013yi,Fischer:2016vfv,Brooks:2020upa},
have the property that each $n$-parton
phase-space point can only be reached via a single shower history. This simplifies the matching procedure in two respects:
Tree-level matrix-element corrections (MEC) for real emission include a single term corresponding
to a unique shower history, at any order.\footnote{The number of contributing shower
histories in standard showers typically grows factorially with the number of emissions.}
Secondly, full phase-space coverage is guaranteed by the addition of direct $2\mapsto n$ branchings~\cite{Hartgring:2013jma,Li:2016yez,Campbell:2021svd},
which exclusively populate ``unordered'' regions of phase space, i.e.~regions that are inaccessible to
nested $2\mapsto 3$ branchings. Matching of sector showers comes at a cost, namely that the ordering/sector constraints
complicate the evaluation of the Born$+1$-jet NLO correction term, which is the main subject of this article.

Here, we derive explicit
expressions for the matching coefficients required for hadronic $Z$ decays, fully
differentially in each of the 2-, 3-, and 4-parton phase spaces. The study case of colour-singlet decay in $e^+e^-$ collisions is especially simple (as we elaborate upon below). We use it as a proof of concept:
we will pay particular attention
to the derivation of the $3$-jet matching coefficient at NLO, and its numerical evaluation
in C++. In \secref{sec:matching-formulae} we introduce our matching scheme,
establish the notation used in this paper, and systematically derive the explicit forms
of the matching coefficients at the 2- (\secref{sec:NNLO2}), and 4-~(\secref{sec:LO4}) and 3-jet (\secref{sec:NLO3}) levels. 
\Secref{sec:sectorsub} describes the subtraction procedure that we implement in order to
evaluate numerically the NLO 3-jet matching coefficient.
In \secref{sec:nF} we provide the $N_F$ contribution, as well as a
consistent treatment of the interference between Born processes with different flavour
contents. For instance, for $Z$ decays at second order,  amplitudes that contain a $Z\to
u\bar{u}$ Born-level coupling and a $g\to d\bar{d}$ vertex must be added to ones with a
$Z\to d\bar{d}$ Born-level coupling and a $g\to u\bar{u}$ vertex; the resulting
(non-singular) interference term in the squared matrix element must be accounted for in
the matched result for it to have the correct fixed-order expansion.
In
\secref{sec:uncertainties}, we give some recommendations on evaluating uncertainties
associated with uncontrolled higher-order terms, and finally we conclude in
\secref{sec:conclusions}. Appendices~\ref{app:Ffunction}--\ref{app:globalsub} include some
auxiliary formulae and alternative prescriptions beyond those covered in the main text, and appendix~\ref{app:matching-alg-descr} lays out our matching procedure in algorithmic form.

\section{Derivation of the matching scheme}
\label{sec:matching-formulae}
In this section, we provide a detailed derivation of a multiplicative matching scheme for hadronic $Z$-boson decays to two jets at NNLO in QCD. The main novelty of this scheme is that the matching conditions are imposed not at the level of a 
(fully or partially) integrated observable --- but fully differentially in each of the $2$-, $3$-, and $4$-parton phase spaces. This ensures that the distribution of \emph{any} IR safe observable will be matched to the stipulated fixed-order accuracy.\footnote{By IR safe we mean both soft and collinear safe;  sometimes also referred to as infrared and collinear (IRC) safe.} To accomplish this, we employ 
{\em multi-differential jet rates} to perform the matching. Deriving the matching conditions in terms of physical cross sections (jet rates), which are guaranteed to be positive definite, also serves to eliminate negative weights, which can have a significant impact on the effective convergence rate. 

Several of the crucial ingredients of this scheme have been developed in previous work over a number of years~\cite{Giele:2011cb,Lopez-Villarejo:2011pwr,Li:2016yez,Campbell:2021svd}. Here, we present, for the first time, the required elements in a single consistent framework, with all expressions worked out in detail.

Starting from the collection of phase-space variables and quantum numbers of an $n$-parton configuration, $\Phi_n$, we denote the resolution measure of the (for now, arbitrary) jet algorithm acting on such a configuration by 
\begin{equation}
t_n = t(\Phi_n)\, ,
\end{equation}
while the clustering prescription (a.k.a.\ clustering map or recombination scheme) of the algorithm is represented by
\begin{equation}
\hat{\Phi}_{n-m}(\Phi_n)\, .
\end{equation}
Note that for parton clusterings with $m \ge 2$ (multi-parton clusterings), we shall restrict our discussion to maps that factorise into simple iterations of single clusterings.

In principle, any (IR safe) jet algorithm can be used to define the resolution measure $t$ and clustering map $\hat{\Phi}$. Since the ARCLUS algorithm~\cite{Lonnblad:1992qd} coincides with the inverse of the $2\mapsto 3$ kinematics map used for gluon emissions in VINCIA, this will be a particularly convenient choice for us. 
The ARCLUS algorithm performs $3 \mapsto 2$ clusterings $\{i,j,k\}\mapsto \{\tilde{ij},\tilde{jk}\}$ according to a resolution variable $t_{[j]/ik}$. To simplify the discussion, let us for now consider the colour-ordered $n$-parton state $\bar{q} g_1 \dots g_{n-2} q$ (we will come back to $g \to q\bar q$ splittings in \secref{sec:bornInterference}). For parton $j$ and its colour neighbours $i=j-1$ and $k=j+1$, ARCLUS computes the following resolution measure
\begin{equation}
t_{[j]/ik}=  \frac{s_{ij} s_{jk}}{s_{ijk}} \,,
\end{equation}
which is known as (the square of) ARIADNE~\cite{Lonnblad:1992tz} $p_\perp$.
The jet resolution is then defined as:
\begin{equation}
\label{eq:tRes}
t_n = \min_{[j]} t_{[j]/ik}(\Phi_n) \,,
\end{equation}
for the smallest value of $t_{[j]}$ in an $n$-parton configuration.\footnote{The sector resolution variable appropriate for $g\to q \bar q$ splittings is discussed in \secref{sec:nF}.} The recombination scheme is dictated by the inverse of any suitable antenna $2\mapsto 3$ map such as those defined in \cite{Lonnblad:1992tz,Kosower:2003bh,Giele:2007di,Gehrmann-DeRidder:2011gkt,Kosower:2022svy}.

In what follows, we will derive the NNLO matching equations for the VINCIA sector shower \cite{Brooks:2020upa}, whose forward map for gluon emissions is an exact inverse of the clustering procedure of the ARCLUS algorithm.

This section is structured as follows. In \secref{sec:fo}, the ARCLUS jet rates are defined at fixed order. Subsequently, the matching of the differential NNLO 2-jet, NLO 3-jet, and LO 4-jet rates is discussed in \secref{sec:NNLO2}, \secref{sec:NLO3}, and \secref{sec:LO4} respectively.

\subsection{Fixed-order differential jet rates}\label{sec:fo}
The ${\cal O}(\alpha_s^2)$ perturbative expansions for the \emph{fully-differential} inclusive 2-, 3-, and 4-jet rates in hadronic $Z$ decays are calculated  as follows (suppressing overall energy-momentum-conserving $\delta$ functions):
\begin{eqnarray}
\mbox{2-Jet Rate~@~\rm{NNLO}:~} \hspace*{-1.7cm} \label{eq:2jetrate} & & \\[1mm]
\frac{\mZ}{8\pi^4} \,\frac{ \d^2 \Gamma}{\d{\Phi}_2} & = & |M_2^0|^2 ~+~
2 \mathrm{Re}[M_2^1 M_2^{0*}] \,+ \int \d \Phi_3 \,|M_3^0|^2 \,\delta^{(2)}\Big({\Phi}_2 - \mapPhi_2(\Phi_3)\Big)\nonumber \\
& & +\, |M_2^1|^2 \,+\, 2\mathrm{Re}[M_2^2M_2^{0*}]\,+ \int\d \Phi_3 \,2\mathrm{Re}[M_3^1M_3^{0*}]\,\delta^{(2)}\Big(\Phi_2 - \mapPhi_2(\Phi_3)\Big) \nonumber \\
& & +\, \int \d \Phi_4 \,|M_4^0|^2 \,\delta^{(2)}\Big({\Phi}_2 - \mapPhi_2(\Phi_4)\Big)~+~{\cal O}(\alpha_s^3)~, \nonumber \\[3mm]
\mbox{3-Jet Rate~@~\rm{NLO}:~}\hspace*{-1.4cm} \label{eq:3jetrate} & & \\[1mm]\frac{\mZ}{8\pi^4} \,\frac{ \d^5 \Gamma}{\d{\Phi}_3} & = & |M_3^0|^2 ~ +~
2 \mathrm{Re}[M_3^1 M_3^{0*}] \,+ \int \d \Phi_4 \,|M_4^0|^2 \,\delta^{(5)}\Big({\Phi}_3 - \mapPhi_3(\Phi_4)\Big)~+~{\cal O}(\alpha_s^3)~, \nonumber \\[3mm]
\mbox{4-Jet Rate~@~\rm{LO}:~}\hspace*{-1.1cm} \label{eq:4jetrate} & & \\[1mm]
\frac{\mZ}{8\pi^4} \,\frac{ \d^8 \Gamma}{\d {\Phi}_{4}} & = & |M_4^0|^2 ~+~{\cal O}(\alpha_s^3)~,\nonumber
\end{eqnarray}
where we have left the dependence of the matrix elements on the respective $n$-parton phase spaces implicit.
Note also that, in addition to the implicit dependence on the phase-space variables, $\Phi_n$, the amplitudes will also in general contain one or more powers of the renormalised coupling evaluated at a specific scale $\mu_M$, 
\begin{equation}
    |M_n|^2 ~\equiv~ |M_n({\Phi}_n;\mu_M)|^2~.
\end{equation}

Eqs.~\eqref{eq:2jetrate}--\eqref{eq:4jetrate} are the expressions we shall seek to match to fixed-order expansions of parton-shower rates below. To make the notation less cumbersome, we adopt a \powheg-inspired notation with $\calB_n$ the lowest-order matrix element for $n$ partons, $\calV_n$ the sum of the integrated real and virtual contributions at NLO, and similarly $\calW_n$ for the NNLO contributions. With this notation, the terms in the above equations are:
\begin{eqnarray}
\calB_2 & = & |M_2^0|^2 \,, \\[2mm]
\calB_3 & = & |M_3^0|^2 \,,\\[2mm]
\calB_4 & = & |M_4^0|^2\,,\\[2mm]
\calV_2 & = & 2 \mathrm{Re}[M_2^1 M_2^{0*}] \,+ \int \d \Phi_3 \,|M_3^0|^2 \,\delta^{(2)}\Big({\Phi}_2 - \mapPhi_2(\Phi_3)\Big) \,, \\[2mm]
\calV_3 & = & 2 \mathrm{Re}[M_3^1 M_3^{0*}] \,+ \int \d \Phi_4 \,|M_4^0|^2 \,\delta^{(5)}\Big({\Phi}_3 - \mapPhi_3(\Phi_4)\Big)\,,\label{eq:V3}\\[2mm]
\calW_2 & = & |M_2^1|^2 \,+\, 2\mathrm{Re}[M_2^2M_2^{0*}]\,+ \int\d \Phi_3 \,2\mathrm{Re}[M_3^1M_3^{0*}]\,\delta^{(2)}\Big(\Phi_2 - \mapPhi_2(\Phi_3)\Big) \nonumber \\
& & +\, \int \d \Phi_4 \,|M_4^0|^2 \,\delta^{(2)}\Big({\Phi}_2 - \mapPhi_2(\Phi_4)\Big)\,.
\end{eqnarray}
We also define a variant of $\calV_3$ in which the 4-parton integral only includes \emph{ordered} clustering sequences, i.e.\ for which the jet clustering measure evaluated on the clustered 3-parton state is larger than that evaluated on the 4-parton state,\footnote{One can choose the jet algorithm such that any unordered 4-parton phase space point (i.e., with \( t_4 > t_3 \)) is not clusterable into a 3-jet state.}
\begin{equation}
\calV_3^{\mathrm O} ~=~ 2 \mathrm{Re}[M_3^1 M_3^{0*}] \,+ \int_{t_3 > t_4} \d \Phi_4 \,|M_4^0|^2 \,\delta^{(5)}\Big({\Phi}_3 - \mapPhi_3(\Phi_4)\Big)\,,
\label{eq:V3O}
\end{equation}
Below, we shall also need the explicit forms of $\calB_2$ and $\calV_2$, the former of which is just the tree-level squared matrix element for $Z$ decay and the latter being the well-known NLO correction~\cite{Ellis:1980wv}, 
\begin{equation}
\label{eq:V2}
\calV_2 ~=~ \frac{3 C_F \alpha_s(\mZ^2)}{4\pi} \calB_2  ~=~ \frac{\alpha_s(\mZ^2)}{\pi} \calB_2 \ \,.
\end{equation}

\subsection{Matching the differential NNLO 2-jet rate}\label{sec:NNLO2}
For a process as simple as $Z$ decay, it is trivial to match a probability-conserving shower to the inclusive NNLO rate, simply by using the full NNLO expression for the total width, $\propto \calB_2 + \calV_2 + \calW_2$, to define a Born-local reweighing factor. 
Since the matching scheme we define preserves this property of strict unitarity, the inclusive NNLO rate is trivially retained.
Nevertheless, we will explicitly show how this works out.

Starting from a standard construction of a unitary shower, see e.g.\ \cite{Giele:2007di,Giele:2011cb,Hartgring:2013jma,Li:2016yez,Campbell:2021svd}, the matching condition for the NNLO inclusive rate is:
\begin{eqnarray}
    \nonumber
    \frac{\mZ}{8\pi^4} \,\frac{ \d^2 \Gamma}{\d{\Phi}_2} & = & \left(\calB_2 + \calV_2 + \calW_2\right) \\
   & = & \left(\calB_2 + \calV_2 + \calW_2\right)\bigg(\Delta_2(m_Z^2,t_c)\,\nonumber  \\ 
    &+& \nonumber \int \,\d\Phi_3 \, \Delta_2(m_Z^2,t_3)\, A_{2\to 3}\, \Delta_3(t_3,t_c)\, \delta^{(2)}(\Phi_2 - \mapPhi_2(\Phi_3)) \\ \nonumber
    &+& \int \,\d\Phi_4\, \Theta(t_3-t_4)\, \Delta_2(m_Z^2,t_3)\,  A_{2\to 3}\, \Delta_3(t_3,t_4)\, A_{3\to 4} \, \delta^{(2)}(\Phi_2 - \mapPhi_2(\Phi_4)) \\
    &+& \int \,\d\Phi_4\, 
    \Theta(t_4-t_3)\, 
    \Delta_2(m_Z^2, t_4)\,
    A_{2\mapsto 4} \, 
    \delta^{(2)}(\Phi_2 - \mapPhi_2(\Phi_4)) \bigg) \, + {\cal O}(\alpha_s^3)\,,
    \label{eq:2jallord}
\end{eqnarray}
where the first line represents simply the fixed-order result and the subsequent lines represent the explicit sum over showered events (with an arbitrary stopping scale $t_c$):
\begin{itemize}
\item The exclusive 2-jet rate $\propto \Delta_2(m_Z^2,t_c)$, the Sudakov factor that represents the no-branching probability of the shower evolution from the 2-parton state. Note that this includes both (iterated) $2\mapsto 3$ and (direct) $2\mapsto 4$ contributions \cite{Li:2016yez}: 
\begin{equation}\label{eq:2ptSud}
\Delta_2  ~=~ \Delta_{2\mapsto 4} \,\Delta_{2\mapsto 3} \, .
\end{equation}
\item The exclusive 3-jet rate which clusters back to the given Born configuration 
\begin{equation}
\propto \Delta_2(m_Z^2,t_3)\Delta_3(t_3,t_c)\delta^{(2)}\big(\Phi_2 - \hat{\Phi}_2(\Phi_3)\big)\,,
\end{equation} where $\Delta_3$ is the $3 \to 4$ Sudakov factor relevant for a $3$-parton state which does not branch further down to the scale $t_c$. It is given by
\begin{equation}\label{eq:3ptSud}
    \Delta_3(t_3,t_c) = \exp\left( - \int^{t_3}_{t_c} \d\Phi_{4} \frac{A_{2\to 4}}{A_{2\to3}} \delta^{(5)}(\Phi_3 - \mapPhi_3(\Phi_4))\right)\, .
\end{equation} 
\item The inclusive rate of 4-jet events that cluster back to the given Born configuration according to an \emph{ordered} sequence of jet clustering scales 
\begin{equation}
\propto \Theta(t_3-t_4)\Delta_2(m_Z^2, t_3)\Delta_3(t_3, t_4)
\delta^{(2)}\big(\Phi_2 - \hat{\Phi}_2(\Phi_4)\big)\,,
\end{equation}
\item The inclusive rate of 4-jet events  that cluster back to the given Born configuration according to an \emph{unordered} sequence of jet clustering scales 
\begin{equation}
    \propto \Theta(t_4-t_3)\Delta_2(m_Z^2, t_4)\delta^{(2)}\big(\Phi_2 - \hat{\Phi}_2(\Phi_4)\big)\,.
\end{equation}
\end{itemize}
A crucial point here is that a unitary shower is \emph{defined} by the fact that the relative rates for $m < n$ exclusive jets and $n$ inclusive jets 
sum to unity. This is true order by order, to all orders, hence even the ${\cal O}(\alpha_s^3)$ term in eq.~\eqref{eq:2jallord} is redundant. 

We remind that we assume the map $\mapPhi_{n-m}(\Phi_n)$ for $m\geq 2$ is nested, {\em viz.} 
\begin{equation}\label{eq:nested}
    \mapPhi_{2}(\Phi_4) = \mapPhi_{2}(\mapPhi_{3}(\Phi_4))\, ,
\end{equation}
which fixes uniquely the {\em intermediate} $3$-parton scale appearing in the last two lines of eq.~\eqref{eq:2jallord}, so that the 3-parton scale, $t_3$, in these lines is indeed identical to that in the second line. (Thanks to the nested antenna map, eq.~\eqref{eq:nested}, this scale is uniquely fixed point-by-point in $\Phi_4$.)

Note also that, for a generic ordered shower, there are parts of the $4$-parton phase space (the ``unordered'' region) that cannot be reached by iterated $2 \mapsto 3$ branchings. The solution implemented in VINCIA is to allow direct $2\mapsto 4$ splittings~\cite{Hartgring:2013jma,Li:2016yez,Campbell:2021svd} to populate those regions. That is what the fourth line in eq.~\eqref{eq:2jallord} represents. 

Since our goal is NNLO matching, we now expand the all-orders 2-particle Sudakov factor up to $\mathcal{O}(\alpha_s^2)$:
\begin{align}
  \Delta_{2\mapsto3}^1(\mZ^2,t_3) &=~ -\int_{t_3}^{\mZ^2} \d\Phi'_{3} \,A^\mathrm{LO}_{2\mapsto3}\, \delta^{(2)}(\Phi_2 - \mapPhi_2(\Phi'_3)) \, ,  \label{eq:sud2to3_1}\\
  \Delta_{2\mapsto3}^2(\mZ^2,t_3) &=~ -\int_{t_3}^{\mZ^2} \d\Phi'_{3} \,A_{2\mapsto3}^{\mathrm{NLO}}\, \delta^{(2)}(\Phi_2 - \mapPhi_2(\Phi'_3)) \, , \\
  \Delta_{2\mapsto4}^2(\mZ^2,t_c) &=~ -\int_{t_c}^{\mZ^2} \d\Phi_{4} \,\Theta(t_4-t_3)\, A_{2\mapsto4}^\mathrm{LO}\, \delta^{(2)}(\Phi_2 - \mapPhi_2(\Phi_4)) \, , 
\end{align}
where the superscripts indicate the order of the expansion in powers of $\alpha_s$.
We shall see that the specific forms of $A_{2\mapsto3}^\mathrm{NLO}$ and $A_{2\mapsto4}^\mathrm{LO}$ are determined by the NLO 3-jet and LO 4-jet rates in sections \ref{sec:NLO3} and \ref{sec:LO4}, respectively.

We also note the LO consistency condition dictated by unitarity of the shower, 
\begin{equation}\label{eq:3to4LO}
    A^\mathrm{LO}_{2\mapsto3} \, A^\mathrm{LO}_{3\mapsto4} = A^\mathrm{LO}_{2\mapsto4} \, .
\end{equation}
\subsection{Matching the differential LO 4-jet rate}\label{sec:LO4}
On the shower side, the inclusive differential rate of $4$-parton configurations receives contributions from ordered $2\mapsto 3$ branchings and unordered $2\mapsto 4$ branchings. To all orders, we then have:
\begin{equation}
\begin{split}
    \frac{\mZ}{8\pi^4} \,\frac{ \d^8 \Gamma}{\d {\Phi}_{4}} = \left(\calB_2 + \calV_2 + \calW_2\right)&\Big[\Delta_{2}(m_Z^2,t_3)A_{2\mapsto3}\Delta_{3}(t_3,t_4)A_{3\mapsto4} \Theta(t_3 - t_4) \\
     &~+~ \Theta(t_4-t_3)\Delta_{2}(m_Z^2,t_4)A_{2\mapsto4}^\mathrm{LO}\Big] \, . 
\end{split}
\end{equation}
Expanding this to ${\cal O}(\alpha_s^2)$ and matching to fixed-order, we find:
\begin{equation}
     \calB_4 \equiv \calB_2\left[\Theta(t_3-t_4)A_{2\mapsto3}^\mathrm{LO}A_{3\mapsto4}^\mathrm{LO} + \Theta(t_4-t_3)A_{2\mapsto4}^\mathrm{LO}\right] ~+~ {\cal O}(\alpha_s^3) \, ,
\end{equation}
and using eq.~\eqref{eq:3to4LO} gives the simple matching condition:
\begin{equation}\label{eq:A2to4}
    A_{2\mapsto4}^\mathrm{LO} = \frac{\calB_4}{\calB_2} + {\cal O}(\alpha_s^3)\, .
\end{equation}
For double-gluon emissions at leading colour in $Z$ decay, this makes the $2\mapsto 4$ emission density identical to the GGG antenna function $A_4^0$~\cite{Gehrmann-DeRidder:2005btv} (modulo the colour and coupling factor). Equivalent functions exist for the $g\to q\bar{q}$ contributions; we defer a detailed discussion of them to sec.~\ref{sec:nF}.
%

\subsection{Matching the differential NLO 3-jet rate}\label{sec:NLO3}
For a parton shower starting from 2-parton configurations normalised to the inclusive NNLO rate, the inclusive differential rate of 3-parton configurations, to all orders, reads:\footnote{This simple form assumes no mixing between different Born states at the 3-parton level. This is true for Z decays but not, e.g., for $H\to Q\bar{Q}g$. In that case, our procedure for treating mixing between different Born states would need to be applied. This is described in \secref{sec:bornInterference}.}
\begin{eqnarray}\label{eq:PS3}
\hspace*{-6cm} & &\nonumber\\ \frac{\mZ}{8\pi^4} \frac{\d^5 \Gamma}{\d{\Phi}_3 } &~=~ &  (\calB_2 + \calV_2 + \calW_2) \Bigg[ \Delta_2(\mZ^2, t_3) 
\, A_{2\mapsto3} \\
& & \qquad\qquad +\, \int\,\d\Phi_{4}\, \Theta(t_4 - t_3) \,\Delta_2(\mZ^2, t_4) \,A_{2\mapsto4}\, \delta^{(5)}({\Phi}_3-\mapPhi_3(\Phi_4))\Bigg]~,\nonumber
\end{eqnarray}
with $A_{2\mapsto3}$ the $2\mapsto 3$ antenna function of the shower (on which we shall impose matching conditions below) and $\Delta_2$ the corresponding no-emission probability, eq.~\eqref{eq:2ptSud}. The first line in eq.~\eqref{eq:PS3} represents the contribution of ordered sequences to the inclusive $3$-jet differential rate, while the second line represents the contribution of unordered clustering sequences. We have again left the phase-space dependence implicit, understanding that terms with a subscript $n$ depend on a corresponding $n$-parton phase space. For example,\footnote{For the specific case of $Z$ decay, the generic notation used here is somewhat overkill; the only explicit dependence on the 2-parton phase space, $\mapPhi_2(\Phi_3)$, is via the total invariant mass, $\mZ^2$.}
\begin{eqnarray}
\calB_2 & \equiv & \calB_2\big( \mapPhi_2( {\Phi}_3 ) \big) \,, \\[2mm]
\Delta_2(\mZ^2, t_3)  & \equiv & \Delta_2\big(\mapPhi_2( {\Phi}_3), \mZ^2, t_3({\Phi}_3) \big)\,,\\[2mm]
A_{2\mapsto3} & \equiv & A_{2\mapsto3}\big( \mapPhi_2(\Phi_3) ,{\Phi}_3 \big) \,. 
\end{eqnarray}

 Expanding eq.~\eqref{eq:PS3} to $\mathcal{O}(\alpha_s)$ and $\mathcal{O}(\alpha_s^2)$ respectively, and comparing to the corresponding (LO and NLO) fixed-order 3-jet rate, we readily obtain for the LO- and NLO-matched $2\mapsto 3$ emission densities:
\begin{eqnarray}
\calB_2 \,A_{2\mapsto3}^\mathrm{LO} & = & \calB_3 + {\cal O}(\alpha_s^2) \,, \label{eq:ALO}\\[2mm]
(\calB_2 + \calV_2) \,\big(1+\Delta_2^1(\mZ^2,t_3)\big)\, A_{2\mapsto3}^\mathrm{NLO} & = & \calB_3 + \calV^{\mathrm O}_3 + {\cal O}(\alpha_s^3) \, ,\label{eq:ANLO}
\end{eqnarray}
with $\calV^{\mathrm O}_3$ given in eq.~\eqref{eq:V3O}. Here we have used the LO $4$-parton matching condition in eq.~\eqref{eq:A2to4}.
Solving eq.~\eqref{eq:ALO} for the antenna function yields the LO matching condition~\cite{Norrbin:2000uu}, 
\begin{eqnarray}
\label{eq:A2to3lo}
A_{2\mapsto3}^\mathrm{LO} & = & \frac{\calB_3}{\calB_2} ~+~ {\cal O}(\alpha_s^2)\,,
\end{eqnarray}
where we note that the ratio $\calB_3/\calB_2$ for $Z$ decay is identical to the GGG antenna function $A_3^0$~\cite{Gehrmann-DeRidder:2005btv} (modulo the colour and coupling factor). This is also the radiation function that would be employed in first-order matrix-element corrections~\cite{Norrbin:2000uu} or \powheg~\cite{Frixione:2007vw} for $Z$ decay. Notice as well that combining eqs.~\eqref{eq:A2to4} and \eqref{eq:A2to3lo} yields the matching condition for $A_{3\mapsto4}^{\rm LO}$:
\begin{eqnarray}\label{eq:A3to4LO} 
    A_{3\mapsto 4}^\mathrm{LO} & = & 
\frac{\calB_4}{\calB_3} ~+~ {\cal O}(\alpha_s^2)\, . 
\end{eqnarray}
Similarly, using eq.~\eqref{eq:ANLO} yields the NLO matching condition for $A_{2\mapsto3}^\mathrm{NLO}$, 
\begin{eqnarray}
A_{2\mapsto3}^\mathrm{NLO} & = & \frac{\calB_3 \,\big(1 - \Delta_2^1(\mZ^2,t_3)\big) + \calV^{\mathrm O}_3}{\calB_2 + \calV_2} \label{eq:A3NLOa}\\[1mm] & = & A_{2\mapsto3}^\mathrm{LO} \left(
1-\Delta_2^1(\mZ^2, t_3) - \frac{\calV_2}{\calB_2} + \frac{\calV^{\mathrm O}_3}{\calB_3}\right)
~+~{\cal O}(\alpha_s^3)\,,\label{eq:A3NLOb}
\end{eqnarray}
where the latter expansion reproduces the form of the matching equation proposed in ref.~\cite{Campbell:2021svd}. The form given in eq.~\eqref{eq:A3NLOa} contains higher-order $\alpha_s$ terms due to the choice of not expanding the $\calB_2+\calV_2$ terms in eq.~\eqref{eq:ANLO}. We return to the impact of such a choice in \secref{sec:uncertainties}.

In the context of NNLO matching to inclusive $Z$ decays, it is this antenna function, $A_{2\mapsto3}^\mathrm{NLO}$, that we shall seek to make fully explicit and implement in the VINCIA sector shower. Few comments are in order:
\begin{itemize}
\item The shower evolution will continue beyond the 4-parton level to generate higher-multiplicity states. However, this further evolution will not affect the ARCLUS jet rates at any order, thanks to the unitarity of the shower and the fact that the ARCLUS algorithm represents the exact inverse of the sector shower. That is one reason why it is convenient to formulate the matching condition in terms of fully differential ARCLUS jet rates in conjunction with sector showers.
\item The explicit calculation of $\calV_3^\mathrm{O}$, for the chosen jet algorithm that defines $\mapPhi_3(\Phi_4)$, is the subject of \secref{sec:sectorsub}. In particular, we show how to arrange for the singularities between real and virtual contributions to cancel analytically, and how to separate remaining terms into ones that can be done analytically (fast) and ones that require numerical treatment (slow). The former, in particular, contain all logarithmic enhancements and dominate $A_{2\mapsto3}^\mathrm{NLO}$.
\item The inclusion of $g\to q\bar{q}$ splittings and interference between different Born processes, for example between $Z\to u\bar{u}\oplus g\mapsto d\bar{d}$ and $Z\to d\bar{d} \oplus g\mapsto u\bar{u}$. These are addressed in \secref{sec:nF}. 
\item Renormalisation-scale and -scheme choices. In the shower, the emission density is  generally evaluated using a scale $\mu_P^2 = {\cal O}({t}_3) < \mu_M^2 = {\cal O}(\mZ^2)$, and typically also employing the so-called CMW scheme~\cite{Catani:1990rr}. This will likewise be relevant in constructing the explicit forms for the matching equations, and is addressed in \secref{sec:muR}. 
\item Higher-order ambiguities. For example, \eqsref{eq:A3NLOa} and \eqref{eq:A3NLOb} are equivalent up to terms of order $\alpha_s^3$; the latter merely makes the modification to the LO antenna function at this order more explicit. In general, any fixed-order matching equation obviously leaves room for higher-order ambiguities, analogous to those between, e.g., MC@NLO and \powheg-style approaches. We return to the explicit choices for these we make below, and in \secref{sec:uncertainties}. 
\end{itemize}

Going back to our central formula eq.~\eqref{eq:A3NLOb}, the expansion of the $2\mapsto 3$ Sudakov factor is given by eq.~\eqref{eq:sud2to3_1} which we rewrite using antenna phase-space factorisation, 
\begin{equation}
\Delta_2^1 = \Delta_{2\mapsto 3}^1(\mZ^2,t_3) ~=~
-\int_{{t}_3}^{\mZ^2} \d\Phi'_{+1} \frac{\calB_3'}{\calB_2} \,,
\label{eq:sudontop}
\end{equation}
where we have used primes to distinguish the 3-parton phase space that is being integrated over from the (fixed) 3-parton configuration that defines $t_3$ (unprimed).
That is, $\calB'_3$ depends on the 2-parton and antenna phase spaces,
\begin{eqnarray}
\calB_3' & \equiv & \calB_3(\Phi'_3) = \calB_3(\Phi_2, \Phi'_{+1})\,,\\[2mm]
\d\Phi_{+1} & = & \frac{1}{16\pi^2 \mZ^2} \d s_{qg}\d s_{g\qbar} \frac{\d \varphi}{2\pi}\,.
\end{eqnarray}
We emphasise that the (massless) antenna phase space is the standard one used, e.g., in refs.~\cite{Gehrmann-DeRidder:2005btv,Giele:2011cb}.

Similarly to other matching schemes, residual power-suppressed contributions impact the final matched result.
%
%
Let us consider a simple example where the matching is performed up to NLO according to the above scheme. In this case, the matrix-element-corrected shower populates the \( q\bar{q}g \) phase space only above the cut-off scale \( t_c \). When computing an IRC-safe observable at \( \mathcal{O}(\alpha_s) \), a residual contribution arises of the form (in the case of \( Z \to \) hadrons):
\[
\int_0^{t_c} d\Phi_3 \, \mathsf{B}_3 \left[O(\Phi_3) - O(\Phi_2)\right],
\]
where \( \mathsf{B}_3 \) denotes the real-emission squared matrix element, and the difference \( O(\Phi_3) - O(\Phi_2) \) vanishes in the soft and collinear limits due to IRC safety.
%
%
At NNLO, the dependence of the matched shower on the cutoff $t_c$ is more intricate. This was discussed in some detail in~\cite{Campbell:2021svd} and we will examine the numerical size of these power-suppressed corrections as part of our future implementation in VINCIA.

\section{Computation of $K^{\rm NLO}$ in VINCIA}
In this section we present the explicit computation of eq.~\eqref{eq:A3NLOb}, and express the result as a (differential) NLO $K$-factor multiplying the LO emission density. The derivation of this factor is complicated by the presence of the sector constraints. We here analyse in detail the implications of these constraints on $K^{\rm NLO}$, relying only on the generic features of the antenna maps. This enables the construction of relatively simple subtraction terms whose integrals reproduce the correct singularity structures. Moreover, in the finite remainder, obtained after combining virtual corrections with the integrated subtraction terms, all logarithmically enhanced contributions are encoded in an analytic function.

We start with the simple integral in eq.~\eqref{eq:sudontop}. This is sometimes referred to as the ``Sudakov on top'', since it originates from the one-loop expansion of the $2 \mapsto 3$ shower evolution~\cite{Hartgring:2013jma}. It is given by:
\begin{eqnarray}
\Delta_2^1 = -\int_{{t}_3}^{\mZ^2} \d\Phi'_{+1} \frac{\calB_3'}{\calB_2} & = & -\frac{\alpha_s C_F}{2\pi}\Bigg[\ln^2{\tau}_3 \, + \, 2\Li(1-{x}_+) \,-\,2\Li({x}_+) \,+\, 3\ln{\tau}_3  \nonumber\\
 && \qquad\qquad+\,6\left(\sqrt{1-4{\tau}_3} - \ln{x}_+ 
 \right) \,-\,2\ln\tau_3\ln{x}_+\Bigg]~\label{eq:topSud}\,,
\end{eqnarray}
with
\begin{equation}
{\tau}_3 = \frac{t_3}{\mZ^2} \label{eq:tau3}\,,\qquad
x_+({\tau}_3) = \frac{1+\sqrt{1-4{\tau}_3}}{2} ~~~\Bigg(\,\to~~ 1 ~\mbox{for}~{\tau}_3 \to 0~\Bigg)\,.
\end{equation}
For reference, we note that the IR limit of eq.~\eqref{eq:topSud} as $\tau_3\to 0$ is:
\begin{equation}\label{eq:Sudontop-expand}
-\frac{\alpha_s C_F}{2\pi}\left[\,\ln^2{\tau}_3 \,+\, 3\ln\tau_3 \,-\, \frac{\pi^2}{3} \,+\, 6\,\right]~.
\end{equation}

If the decaying particle is a scalar instead of a vector, then the 3-parton matrix element includes an additional non-singular term $\alpha_s C_F/(\pi m_H^2)$, which implies a further term to be added to eq.~\eqref{eq:topSud} in the case of Higgs decay,
\begin{equation}
-\frac{\alpha_s C_F}{2\pi}\Big[\,
\sqrt{1-4\tau_3} \,+\, 2\tau_3 \big(\ln (x_+) \,-\, \ln(1-x_+) \,\big)
\Big]~\label{eq:topSudH}
\end{equation}
whose IR limit is $-\alpha_s C_F/(2\pi)$.

Eq.~\eqref{eq:V3O} contains two terms. 
In the case of $Z$ decays, we will deal with the first one analytically, using dimensional regularisation and casting it in a way that makes its pole structure and finite remainders explicit. We wish to highlight that the one-loop correction could in principle also be taken from a (numerical) one-loop provider.
Our main remaining task is then to bring the second term, representing the integral over singly-clustered 4-parton configurations, to an equivalent form --- with explicit poles that cancel those of the first term, and (numerically stable) finite remainders. 

We focus on the integral over 4-parton states that cluster back to the 3-parton state at hand,
\begin{equation}
I_{3\mapsfrom 4}(\Phi_3) = \int_{t_3>t_4}\d\Phi_4\, \calB_4\, \delta^{(5)}\Bigg(\Phi_3 - \hat{\Phi}_3(\Phi_4)\Bigg)~,\label{eq:I43}
\end{equation}
where we reiterate that the restriction to ordered clusterings here, ${t}_3>t_4$, came from matching the fixed-order integral (which included all of phase space) to a shower that includes direct ${\cal{O}}(\alpha_s^2)$ $2\mapsto 4$ branchings in the unordered region.\footnote{Note that exactly the same matching equation would have resulted, with the same form of eq.~\eqref{eq:I43}, should we have used a modified form of the jet algorithm in which direct $4\mapsto 2$ clusterings would be allowed.}

We start by setting up the notation following the conventions of ref.~\cite{Gehrmann-DeRidder:2005btv}. The four-parton final state $(qgg\bar{q})$ is labelled as follows
\begin{align}
(1_q,3_g,4_g,2_{\Bar{q}}) \, ,
\end{align}
while the inverse mapping $3\mapsfrom 4$ leads to a $qg\Bar{q}$ state denoted as follows
\begin{align}
(\hat{1}_q,\hat{3}_g,\hat{2}_{\Bar{q}}) \, .
\end{align}
The fact that we are using a deterministic sequential-clustering jet algorithm divides the phase space up into sectors according to the gluon that has the lowest value of the jet resolution measure. 
At the colour-ordered level, there are then two sectors, which we label $A$ and $B$, corresponding to clusterings of gluons $3$ and $4$ respectively. Thus,
\begin{equation}
I_{3 \mapsfrom 4}(\Phi_3) = \calB_2
\int_{\hat t_3>t_{4A}} \d\Phi_{+1}^A \Theta\big( t_{4B} - t_{4A}\big) A^\mathrm{LO}_{2\mapsto 4} 
~+~ \left(A\leftrightarrow B\right)~,\label{eq:I43ab}
\end{equation}
again with the standard massless 
antenna phase space (in $4-2\epsilon$ dimensions)
\begin{eqnarray}\label{eq:PS-eps}
\d\Phi_{+1}^A & = & \left(\frac{4\pi \mu^2}{\hat{s}_{qg}}\right)^\epsilon \frac{1}{\Gamma(1-\epsilon)}\frac{\hat{s}_{qg}}{16\pi^2}\, \d y_{13} \d y_{34} \d y_{14} \left(y_{13} y_{34} y_{14}\right)^{-\epsilon} \frac{\d\Omega_{2-2\epsilon}}{\Omega_{2-2\epsilon}}\label{eq:antPS} \\[3mm]
& \stackrel{\epsilon \to 0}{=} & 
\frac{\hat{s}_{qg}}{16\pi^2} \d y_{13}\d y_{34} \frac{\d \varphi}{2\pi}\,, \label{eq:antPS4D}
\end{eqnarray}
where the scaled invariants are $y_{ij} = s_{ij}/\hat{s}_{qg}$ (with $y_{13}+y_{14}+y_{34} = 1$), and the three- and four-parton resolution scales defined by our jet-clustering measure are 
\begin{eqnarray}
\hat t_3 & = & \frac{\hat s_{qg}\hat s_{g\qbar}}{\mZ^2}\,,\label{eq:t3}\\[2mm]
t_{4A} & = & \frac{s_{13}s_{34}}{s_{134}}\,, \label{eq:t4A}\\[2mm]
t_{4B} & = & \frac{s_{24}s_{34}}{s_{234}}\,.
\label{eq:t4B}
\end{eqnarray}

The fact that the step functions in eq.~\eqref{eq:I43ab} involve the invariants  $s_{24}$ and $s_{234}$, which go outside of the local $2\mapsto 3$ $\d\Phi_{+1}^A$ and are only determined by the explicit kinematic map (i.e., the inverse of the clustering map), is the main challenge in working with these integrals. We have considered two possible paths to dealing with this, one based on global antenna subtraction, and one based on a new sector antenna subtraction.

In both cases, the ultimate goal is to construct a numerical contribution which is stable and quick to evaluate, while achieving pole cancellation analytically.
The formulations presented here are quite general and work for any antenna function that appears in the integrand, given properly constructed subtraction terms in $4-2\epsilon$ dimensions. 

\subsection{NLO 3-jet matching with sector subtraction}
\label{sec:sectorsub}

Here, we present a computation of the leading-colour (LC) NLO 3-jet matching coefficient, in which the subtraction term is also sectorised. An alternative approach, derived via global antenna subtraction, is given in appendix~\ref{app:globalsub}.

Without loss of generality, we focus on the $A$ sector and then later show how to bootstrap the result for the $B$ sector, namely we wish to compute
\begin{equation}\label{eq:IAsector}
I^{A}_{3\mapsfrom 4} =\calB_3
\frac{g^2 N_C}{A_3^0(\hat{1}_q,\hat{3}_g,\hat{2}_{\Bar{q}})}\int_{\hat{t}_3>t_{4A}} \d\Phi_{+1}^A \Theta\big( t_{4B} - t_{4A}\big) A_4^0 \,,
\end{equation}
where we have now made colour and coupling factors explicit, multiplying the GGG $A_3^0$ and $A_4^0$ antenna functions for $Z$ decay~\cite{Gehrmann-DeRidder:2005btv}. The three- and four-parton resolution scales, ${t}_i$, and the antenna phase space are defined in \eqsref{eq:antPS} -- \eqref{eq:t4B}.

\begin{figure}[t]
\centering
\includegraphics*[width=0.5\textwidth]{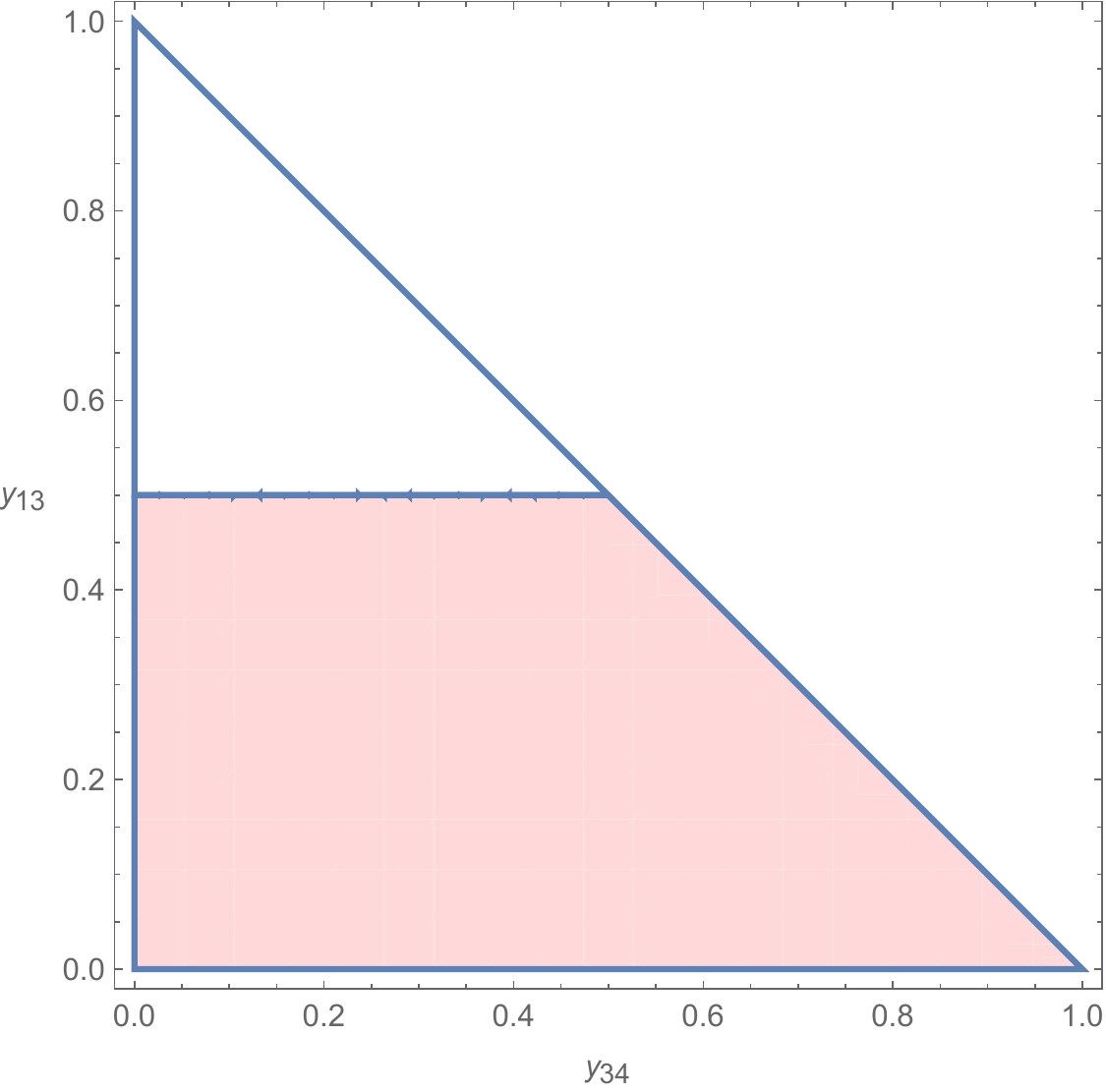}
\caption{The Dalitz plane for the evaluation of eq.~\eqref{eq:IAsector}. All singularities persist in the red-filled polygon, while the empty triangle is singularity-free by virtue of the sector constraint.} 
\label{fig:Dalitz}
\end{figure}

To understand how to best perform the computation, it is easiest to investigate the sector constraint in eq.~\eqref{eq:IAsector} upon implementing an explicit kinematic map. The details of the derivation can be found in appendix~\ref{app:secanal}.
The key result of the analysis is illustrated in fig.~\ref{fig:Dalitz}, which highlights that the sector constraint forces the singularities to lie in the red-shaded region of the radiation phase space.
%
Crucially, this analysis holds for any local kinematic map.
Moreover, we can construct a subtraction term that is only nonzero for $y_{13} < \frac12$, i.e.~in the red region of fig.~\ref{fig:Dalitz}. 


A suitable subtraction term must incorporate the correct singular limits of $A_4^0$ \textit{only} in the red region. In particular the subtraction term must reproduce the full $g\to gg$ splitting function in the $3 || 4$ limit.
With this understanding we introduce the following function in the 4-parton phase space\footnote{We note that improved antenna subtraction terms have also been discussed in ref.~\cite{Braun-White:2023sgd}.}
\begin{align}\label{eq:D30def}
    \Ds(1_q,3_g,4_g) = \frac{1}{s_{134}} \left( \frac{2 y_{14}}{y_{13} y_{34}} + \frac{2y_{13}}{(1-y_{13}) y_{34}} +(1-\epsilon) \frac{y_{34}}{y_{13}} + \frac{2 y_{13} y_{14}}{y_{34}}\right) \, ,
\end{align}
where $y_{14} = 1- y_{13} - y_{34}$.
%
Indeed this subtraction function has all the correct limits:
\begin{equation}
    \lim_{y_{13} \to 0} \Ds = \frac{1}{y_{13}} p_{gq}(z,\epsilon) \, ,\quad
    \lim_{y_{34} \to 0} \Ds = \frac{1}{y_{34}} p_{gg}(z,\epsilon) \, ,
\end{equation}
where $z$ is the energy (light-cone momentum) fraction carried by $3_g$ and the DGLAP splitting functions are
\begin{align}\label{eq:pgg&pgq}
    p_{gq}(z,\epsilon) = \frac{1+(1-z)^2}{z} - \epsilon z \, , \quad p_{gg}(z,\epsilon) = 2\frac{(1-z(1-z))^2}{z(1-z)} \, .
\end{align}

We can now utilise the subtraction function $\Ds$ in eq.~\eqref{eq:D30def} to rewrite the core eq.~\eqref{eq:IAsector} as: 
\begin{multline}
\label{eq:RR-sub}
I_{3\mapsfrom 4}^A =
\,\calB_3
\frac{g^2 N_C}{A_3^0(\hat{1}_q,\hat{3}_g,\hat{2}_{\Bar{q}})}\, \int_{\hat{t}_3>t_{4A}} \d\Phi_{+1}^A \Theta\big( t_{4B} - t_{4A}\big) \bigg[ A_4^0 - \,A_3^0(\hat{1}_q,\hat{3}_g,\hat{2}_{\Bar{q}})\, \Ds \,\Theta\bigg(y_{13} < \frac12\bigg) \bigg] \\
+ \calB_3 \, g^2 N_C \, \int_{\hat{t}_3>t_{4A}} \d\Phi_{+1}^A \Theta\big( t_{4B} - t_{4A}\big) \bigg[ \Ds \,\Theta\bigg(y_{13} < \frac12\bigg) \bigg] \, .
\end{multline}
The first line is free of singularities and can be integrated in 4 dimensions via a MC procedure. The integral of the subtraction term is a bit more complicated than usual, due to the presence of the sector constraint. 

To confirm that we have a proper subtraction term, we integrate the second line of eq.~\eqref{eq:RR-sub} over the antenna phase space, and check that we recover the pole structure of the real-virtual term.
This gives\footnote{To obtain this expression, charge renormalisation has been effected, i.e. $\alpha_s(\mu_R) \to \alpha_s(\mu_R) (4\pi)^{-\epsilon} e^{\epsilon \gamma_E}$.}
\begin{align}\label{eq:D30wosector}
    \nonumber
    (8\pi^2) \, \int_{\hat{t}_3>t_{4A}}&\d\Phi_{+1}^A \, \Ds \,\Theta\bigg(y_{13} < \frac12\bigg)  \\
    &= \left(\frac{\mu^2_R}{\mZ^2}\right)^{\epsilon} \frac{\hat{y}_{qg}^{-\epsilon} }{2} \left(\frac{2}{\epsilon^2} + \frac{10}{3\epsilon} + \frac{453}{48} - \pi^2 - \frac{\ln 2}{3}  \right) - \mathcal{F}(\hat{y}_{g\Bar{q}}) \, ,
\end{align}
where $\mathcal{F}(\hat{y}_{g\Bar{q}})$ is a finite function which is given in appendix~\ref{app:Ffunction}. 

This is indeed the correct pole structure required to cancel out the poles in the real-virtual term, see appendix~\ref{app:virtuals}. The canonical GGG sub-antenna, i.e.~$d_3^A$, gives {\em half} of the $g\to gg$ splitting function in the $y_{34} \to 0$ limit. In this case, integrating over the whole Dalitz triangle results in the correct single pole, i.e. $10/3\epsilon$. With our new subtraction term given in eq.~\eqref{eq:D30def}, although we recover the full $p_{gg}$ in the $y_{34} \to 0$ collinear limit, we still obtain the correct single pole because we restrict the integration domain to $y_{13}<\frac12$.

However, the second line of eq.~\eqref{eq:RR-sub} requires that we implement the full sector constraint, $\Theta(t_{4B}-t_{4A})$. This does not present any difficulty, because as we see in eq.~\eqref{eq:D30wosector} all the poles are correctly reproduced. Therefore, in addition to eq.~\eqref{eq:D30wosector} we have the following left-over term:
\begin{align}\label{eq:leftover}
\int_{\hat{t}_3>t_{4A}}\d\Phi_{+1}^A \,& \Ds \,\Theta\bigg(y_{13} < \frac12\bigg)  \bigg[\Theta\big( t_{4B} - t_{4A}\big) - 1\bigg] \, ,
\end{align}
which is integrable in four dimensions.

\subsubsection{Summary of matching by sector subtraction}
\begin{figure}[t]
\centering
\includegraphics*[width=0.49\textwidth,page=3]{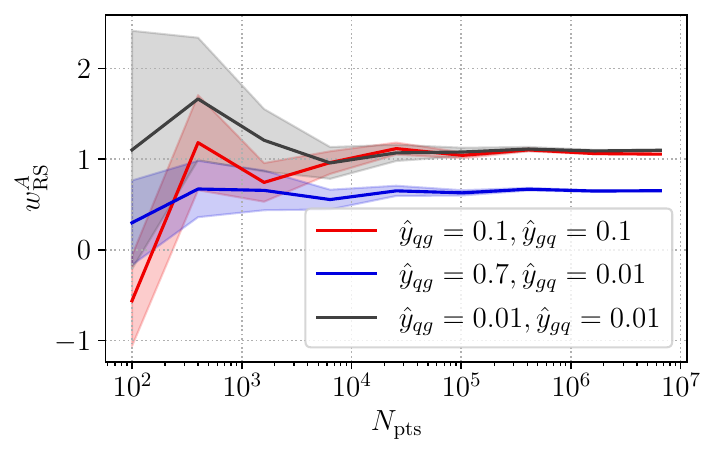}
\includegraphics*[width=0.49\textwidth,page=4]{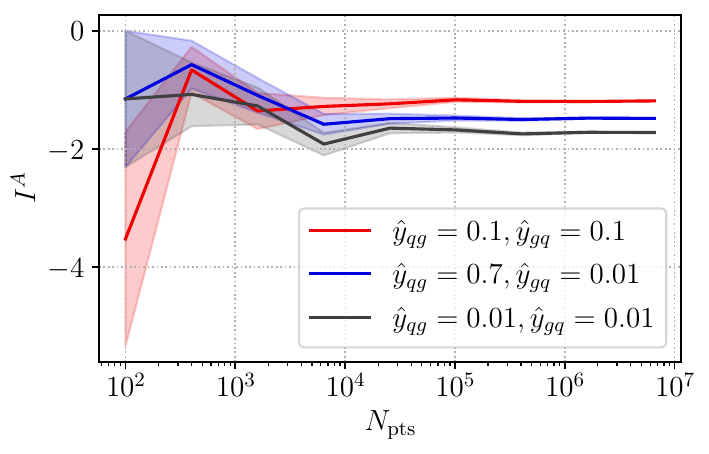}
\caption{The numerical evaluation of the $w_{\rm{RS}}^A$ (left) and $\mathcal{I}^A$ (right) contributions for various Born+1 configurations, as a function of the number of points used in the MC evaluation.
}
\label{fig:wRS-secA}
\end{figure}

We are in a position to collate all the pieces to rewrite eq.~\eqref{eq:IAsector}, which gives
\begin{align}\label{eq:RR-sub-final-A}
\nonumber
I_{3\mapsfrom 4}^A &= \calB_3
 \left(\frac{\alpha_s N_C}{2\pi}\right)\left(\frac{\mu^2_R}{\mZ^2}\right)^{\epsilon}\, \left(\frac{\hat{y}_{qg}^{-\epsilon} }{2} \left(\frac{2}{\epsilon^2} + \frac{10}{3\epsilon} + \frac{453}{48} - \pi^2 - \frac{\ln 2}{3}  \right) - \mathcal{F}(\hat{y}_{g\Bar{q}}) \right) \\
&+\calB_3 \left(\frac{\alpha_s N_C}{2\pi}\right)\, \left( w^A_{\rm RS}(\hat{y}_{qg},\hat{y}_{g\Bar{q}})+ \mathcal{I}^A(\hat{y}_{qg},\hat{y}_{g\Bar{q}}) \right) \, ,
\end{align}
where we defined:
\begin{align}\label{eq:wRS}
    w^A_{\rm RS} =\frac{8\pi^2}{A_3^0(\hat{1}_q,\hat{3}_g,\hat{2}_{\Bar{q}})}\int_{\hat{t}_3>t_{4A}} \d\Phi_{+1}^A \Theta\big( t_{4B} - t_{4A}\big) \bigg[ A_4^0 - \, A_3^0(\hat{1}_q,\hat{3}_g,\hat{2}_{\Bar{q}})\, \Ds \,\Theta\bigg(y_{13} < \frac12\bigg) \bigg] \, ,
\end{align}
and 
\begin{equation}\label{eq:Ifunction}
\mathcal{I}^A = 8 \pi^2 \int_{\hat{t}_3>t_{4A}}\d\Phi_{+1}^A \, \Ds \,\Theta\bigg(y_{13} < \frac12\bigg)  \bigg[\Theta\big( t_{4B} - t_{4A}\big) - 1\bigg] \, .
\end{equation}
The integrals in eqs.~\eqref{eq:wRS} and~\eqref{eq:Ifunction} are computed numerically and shown in fig.~\ref{fig:wRS-secA} for various Born+1 configurations. We notice good convergence with increasing the number of sampling points.
In the MC evaluation of fig.~\ref{fig:wRS-secA} we use the sampling of the antenna phase space $\d\Phi_{+1}$  described in ref.~\cite{Weinzierl:1999yf}. These plots are generated with the default antenna map used in VINCIA \cite{Kosower:2003bh}.

\subsubsection{Bootstrapping the $B$ sector}

\begin{figure}[th]
\centering
\includegraphics*[width=0.49\textwidth,page=4]{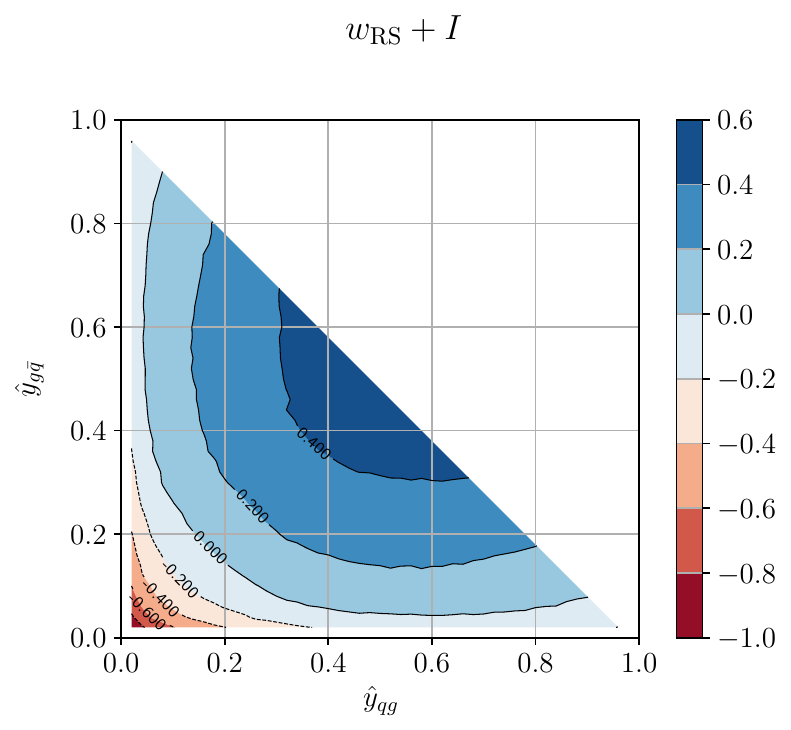}
\includegraphics*[width=0.49\textwidth,page=3]{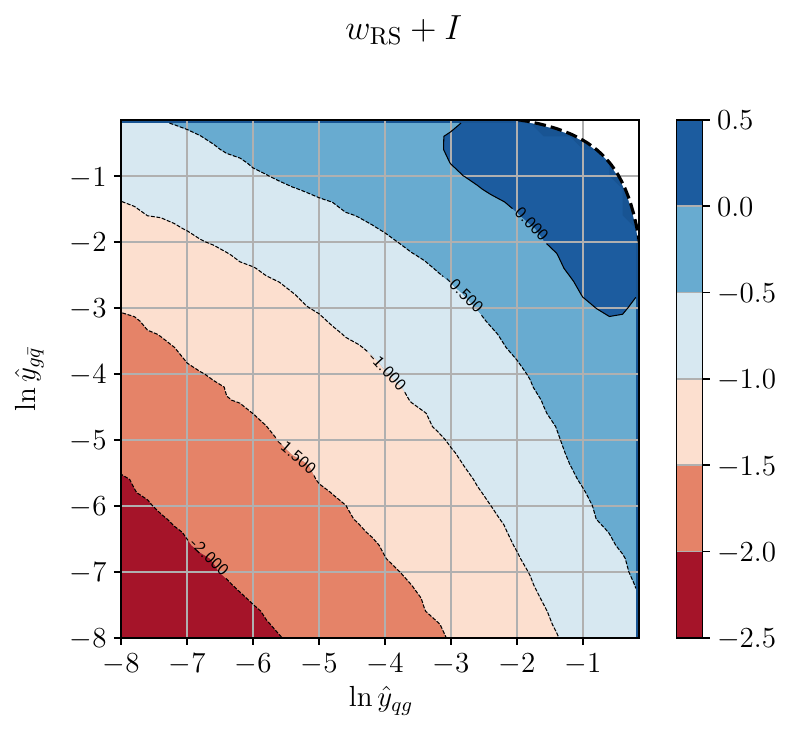}
\caption{The numerical contribution from eq.~\eqref{eq:wRS-I-tot} as a function of $(\hat{y}_{qg}, \hat{y}_{g\bar q})$ (left) and in logarithmic space (right).
}
\label{fig:wRS-I-hm}
\end{figure}

To obtain the final result as given by eq.~\eqref{eq:I43ab}, we must include the contribution from the $B$ sector, i.e. the term $A \leftrightarrow B$. This does not require any further computation, as the result can be obtained from the $A$ sector by the simple swap
\begin{align}
\hat{y}_{qg} \to \hat{y}_{g\bar{q}}, \quad \hat{y}_{g\bar{q}} \to \hat{y}_{qg} \, ,
\end{align}
and thus
\begin{align}\label{eq:RR-sub-final-B}
\nonumber
I_{3\mapsfrom 4}^B(\hat{y}_{qg},\hat{y}_{g\bar{q}}) &= I_{3\mapsfrom 4}^A(\hat{y}_{g\Bar{q}},\hat{y}_{qg}) \,.
\end{align}
The numerical contribution from both sectors,
\begin{equation}
\label{eq:wRS-I-tot}
\left(\frac{\alpha_s N_C}{2\pi}\right) \left( w^A_{\rm RS}+ \mathcal{I}^A + w^B_{\rm RS}+ \mathcal{I}^B\right) \, ,
\end{equation}
is shown in fig.~\ref{fig:wRS-I-hm} as a function of the Born+1 variables.

\subsubsection{The $\KNLO$ factor}
Adding in the virtual corrections, which are given in appendix~\ref{app:virtuals}, we are now ready to write down the final result of eq.~\eqref{eq:V3O},
\begin{align}
\label{eq:V3Ofinal}
    \calV_3^{\rm O}  = \calB_3 \times \KNLO(\hat{y}_{qg},\hat{y}_{g\bar{q}})\, ,
\end{align}
where the $N_C$ piece is given by
\begin{align}\label{eq:KNLO}
\KNLO_{N_C} = \left(\frac{\alpha_s N_C}{2\pi}\right) \left( w^A_{\rm RS}+ \mathcal{I}^A + w^B_{\rm RS}+ \mathcal{I}^B + w_{\mathcal{V}}\right) \, ,
\end{align}
and $w_{\mathcal{V}}$ is the finite analytic function that results from combining virtual corrections and the integrated subtraction term, i.e. the first line of eq.~\eqref{eq:RR-sub-final-A} and its equivalent for sector $B$. The expression of the latter is somewhat lengthy and thus we do not report it here. Nevertheless, it is instructive to extract the logarithmic structure present in the analytic function $w_{\mathcal{V}}$ in all unresolved limits of the Born+1 state. Using the notation of eq.~\eqref{eq:tau3}, we obtain ($\tau_3 = \yqg \yqbg$)
\begin{align}\label{eq:wNC-expand}
    \lim_{\tau_3 \to 0} w_{\mathcal{V}} \to -\frac12 \ln^2 \tau_3 - \frac{10}{3} \ln \tau_3 - \frac{11 \ln2}{6} + \frac{\pi^2}3 -\frac{31}{72} \, .
\end{align}

\begin{figure}[th]
\centering
\includegraphics*[width=0.49\textwidth,page=2]{figures/plot-Zqq-2d-NC.pdf}
\includegraphics*[width=0.49\textwidth,page=1]{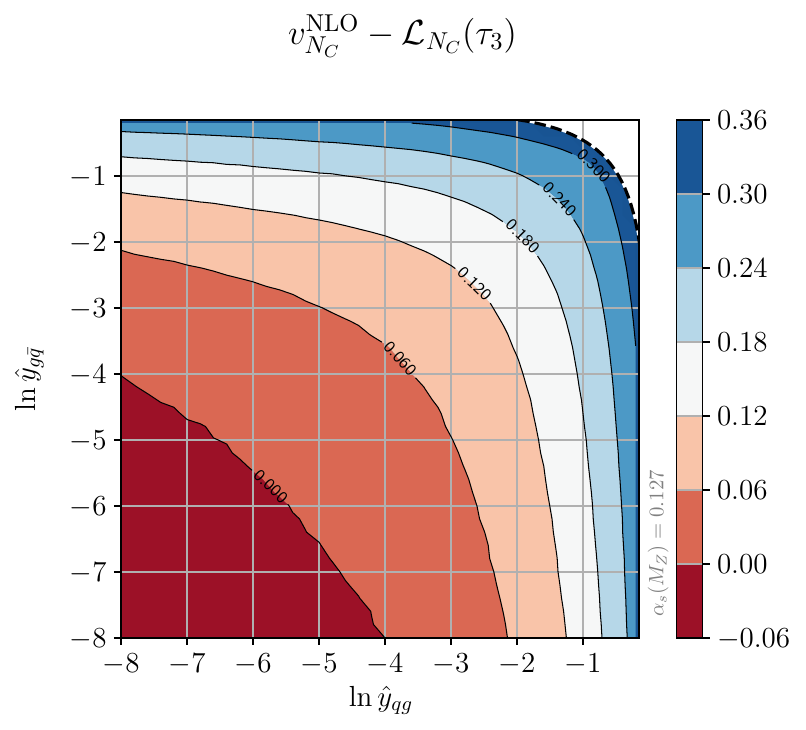}
\caption{The contribution $\KNLO_{N_C}$ as a function of $(\hat{y}_{qg}, \hat{y}_{g\bar q})$ (left) and in logarithmic space (right). Here, we subtract the expected logarithmic terms $\mathcal L(\tau_3) = \frac{\alpha_s N_C}{2\pi} \left(-\frac12 \ln^2 \tau_3 - \frac{10}{3} \ln \tau_3\right)$, with $\tau_3 = \hat y_{qg} \hat y_{g\bar q}$ in eq.~\eqref{eq:wNC-expand} from the full correction.
}
\label{fig:KNLO-sub-hm}
\end{figure}

\subsection{The $N_F$ piece}
\label{sec:nF}

For the gluon-splitting contributions to the 4-parton integrals, there are two further subtleties:
\begin{figure}[t]
\centering
\includegraphics*[width=0.52\textwidth]{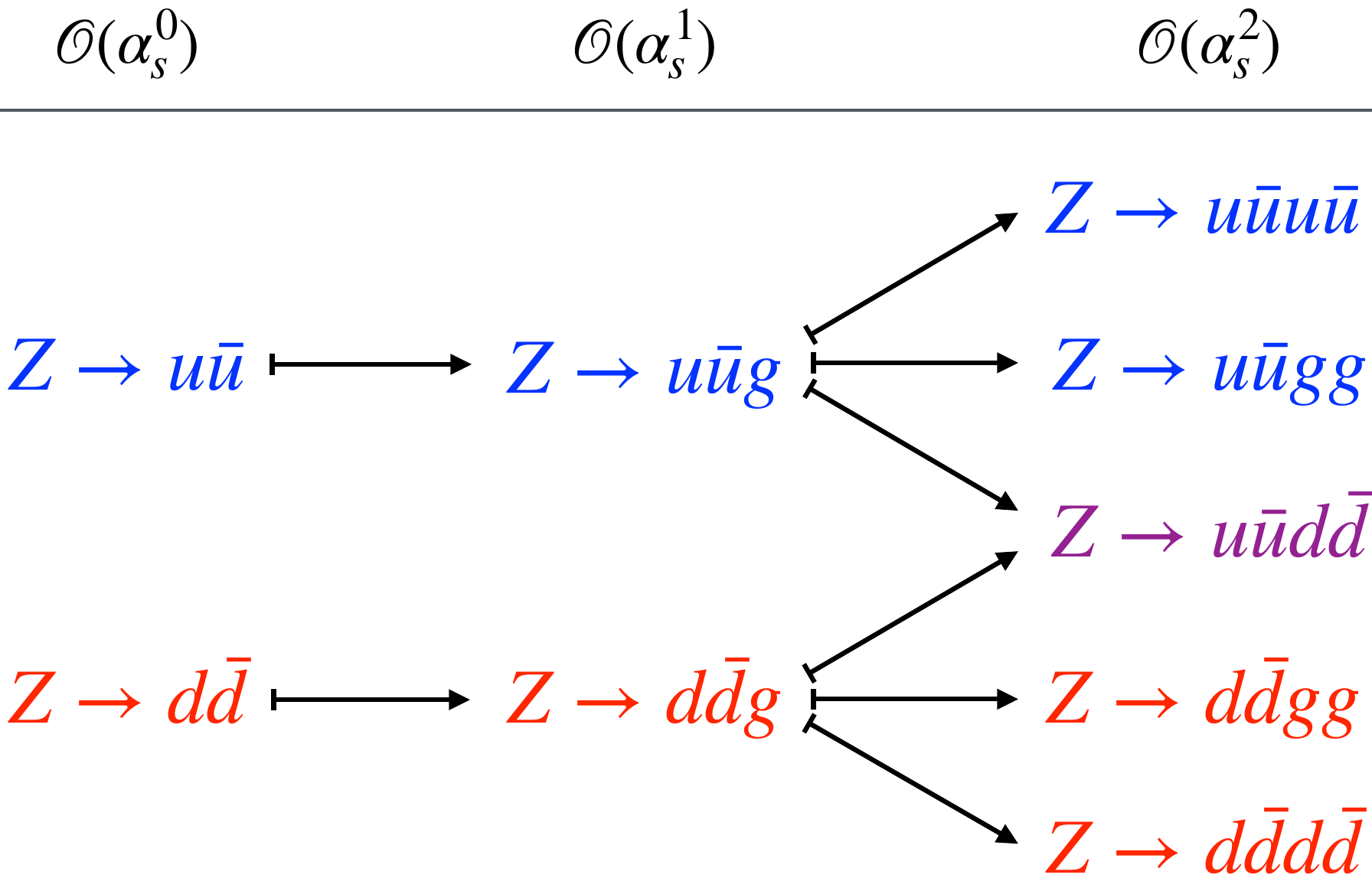}
\caption{Illustration of shower histories starting from $Z\to u\bar{u}$ (blue) and $Z\to d\bar{d}$ (red). Both histories are allowed to contribute to the \emph{same} phase-space points for $Z\to u\bar{u}d\bar{d}$ (purple). \label{fig:4qborns}}
\end{figure}
\begin{enumerate}
\item Interference between shower histories that originate from different Born processes: e.g., $Z\to u\bar{u}d\bar{d}$ receives contributions from both $Z\to u\bar{u}g \,\otimes\, g\to d\bar{d}$ and from $Z\to d\bar{d}g \,\otimes\, g\to u\bar{u}$ shower histories. This is illustrated in \figref{fig:4qborns}. 
Each of these effectively represents an integration channel with a specific singularity structure. After matching, these need to sum up to the full four-parton squared matrix element including interference terms, point-by-point in phase space and with the correct combinations of $q\bar{q}Z$ couplings. We note that, in fixed-order contexts, this subtlety can often be neglected, as fermion antisymmetry implies that the interference cancels in observables that are not sensitive to charge signs. Nevertheless this would not be true of the subsequent shower evolution, and moreover we shall want the matched MC evolution algorithm to be NNLO correct not only for charge-blind observables but also for charge-sensitive ones. 
\item The sector resolution variable we choose for $g\to q\bar{q}$ splittings has to be slightly different from the evolution variable $t_{[j]/ik}$ that was used in the gluon-emission case~\cite{Lopez-Villarejo:2011pwr,Brooks:2020upa}. (Otherwise neighbouring collinear $gq$ singularities would be undercounted.)
\end{enumerate}

We shall now address each of these subtleties in turn, before presenting the full form of the $N_F$ part of $\calV^{\rm O}_3$.

\subsubsection{Multiple interfering Born states}
\label{sec:bornInterference}

Consider, e.g., $Z\to u\bar{u}d\bar{d}$. In a global antenna shower, the total rate in each $u\bar{u}d\bar{d}$ phase-space point would result from a sum over four different histories:
\begin{multicols}{2}
\begin{enumerate}
\item $u\bar{u} \mapsto (ug\mapsto u\bar{d}d)\bar{u}\,$,
\item $u\bar{u} \mapsto u(g\bar{u}\mapsto \bar{d}d\bar{u})\,$,
\end{enumerate}
\columnbreak
\begin{enumerate}  \setcounter{enumi}{2}
\item $d\bar{d} \mapsto (dg\mapsto d\bar{u}u)\bar{d}\,$,
\item $d\bar{d} \mapsto d(g\bar{d}\mapsto \bar{u}u\bar{d})\,$.
  \end{enumerate}
\end{multicols}  
In a sector shower, one has a choice: sectorise across all four of these histories so that only \emph{one} of them is allowed to contribute to each phase-space point, or sectorise only across subsets that share the same Born configuration. For hadronic $Z$-decays, this choice is not critical since the Born-level couplings to different quark flavours have similar orders of magnitude. But to preempt cases like $H\to q\bar{q}$ for which the Born-level couplings can be vastly different, we opt for the latter, namely to sectorise only between shower histories that share the same Born-level configurations. This is in line with what is already done in VINCIA's sector merging~\cite{Brooks:2020mab}.

This implies that, during the shower evolution off a $u\bar{u}$ Born state, only sectors that represent clusterings back to configurations with at least one $u\bar{u}$ pair will be taken into consideration (for the purpose of sectoring), while sector conditions are not checked for clusterings that represent other Born states.  Similarly for the shower evolution off $d\bar{d}$ states, no sector conditions are imposed for clusterings that correspond to $u\bar{u}$ Born states (and analogously for higher numbers of flavours). In this sense, partons that are
present at the Born level are also considered ``hard'' or ``unclusterable''.

Since there is no sectoring between different Born states, the sector-shower rate for parton states that can be reached from multiple Born states will thus contain a sum of terms, one for each distinct Born state. Each such term  mimics the leading singularity structure of a specific subset of squared amplitudes --- involving that particular set of Born couplings.\footnote{
Of course the full matrix elements will include interferences between the different Born processes. 
In the context of matching, it is important to note that it is the \emph{sum} over sector-shower histories that we shall seek to match to the full result. Below, we introduce a few extra notational devices to hopefully make it clear when  and where interferences are included.}

\begin{figure}[t]
\centering
\includegraphics*[width=0.95\textwidth]{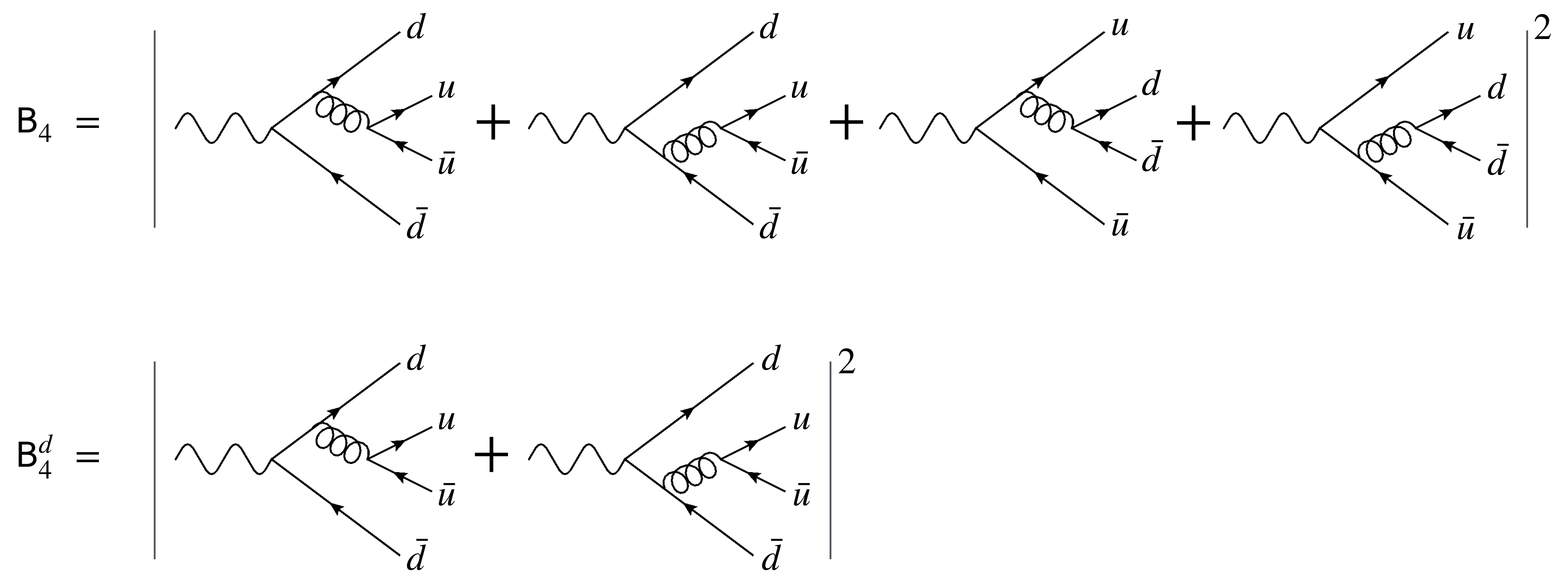}
\caption{Illustration of multiple Born histories, for a $Z\to u\bar{u}d\bar{d}$ final state. {\sl Top:} the full squared matrix element, denoted $\calB_4$, contains an interference term between  diagrams with $\bar{d}dZ$ couplings and diagrams with $\bar{u}uZ$ ones. 
{\sl Bottom:} in the coefficient denoted $\calB_4^d$, only diagrams with Born-level $\bar{d}dZ$ coupling are included. 
\label{fig:Born-diagrams}
}
\end{figure}

Continuing with the example of the 4-parton state $Z\to u\bar u d \bar d$, we need the sum of the two contributions, 
\begin{equation}
Z\to u\bar{u}g \otimes g\mapsto d\bar{d} ~~~+~~~ Z \to d\bar{d}g \otimes g \mapsto u\bar{u}\,,
\end{equation}
to reproduce the full matrix element. Fig.~\ref{fig:Born-diagrams} illustrates the full squared matrix element for the 4-parton state (top row), as well as the contribution proportional to a single $Z$ coupling (bottom row), which we will use in constructing the interference-correction factor below.

For the path $Z \to d \bar d \oplus g\mapsto u\bar{u}$, we generalise eqs.~\eqref{eq:A2to4} and \eqref{eq:A3to4LO} to include an interference-factor, $\bb{k_I}$:
\begin{eqnarray}
A_{3\mapsto 4}^\mathrm{LO} & = &
\bb{k_I^{du}}\,\frac{  \calB_4^d}{\calB_3^d} ~+~ {\cal O}(\alpha_s^2)\,,\label{eq:A3to4LOkI}\\
A_{2\mapsto 4}^\mathrm{LO} & = &\bb{k_I^{du}}\, 
\frac{ \calB_4^d}{\calB_2^d} ~+~ {\cal O}(\alpha_s^3)~, \label{eq:A2to4LOkI}
\end{eqnarray}
where $\calB_{2,3}^q$ are the Born-level squared matrix elements for $Z\to q \bar q$ and $Z\to q\bar{q}g$ respectively, and 
\begin{equation}
    \calB_4^d = \calB_2^d \,(2g^4 C_F) \,B_4^0(1_d, 3_{u}, 4_{\bar u}, 2_{\bar d})\,,\qquad \calB_4^u = \calB_2^u \,(2g^4 C_F) \, B_4^0(1_u, 3_{d}, 4_{\bar d}, 2_{\bar u})\,,
\end{equation}
represent squared matrix elements involving only a single Born coupling. 
The term $k_I$ partitions the full matrix-element correction, including the (non-singular) interference terms, over both Born states,
\begin{equation}
\label{eq:Born-kI}
    \bb{k_I^{du}}~=~ \frac{\calB_4}{\calB_4^d + \calB_4^u} = 1+ \frac{\calB_4^{du}}{\calB_4^d + \calB_4^u}\,,
\end{equation}
where $\calB_4^{du}$ is the pure interference term in the squared matrix element appearing in the top line of fig.~\ref{fig:Born-diagrams}. This term is proportional to the mixed $Z$-quark couplings and to the four-quark antenna function dubbed $\hat B_4^0$ of ref.~\cite{Gehrmann-DeRidder:2005btv}.

For same-flavour $Z\to q \bar q q \bar q$, we use the identical-flavour four-parton matrix element in the definition of the interference factor,
\begin{equation}
\begin{split}
\bb{k_I^{qq}} &= \frac{1}{B_4^0(1_q,3_q,4_{\qbar},2_{\qbar}) + B_4^0(3_q,1_q,2_{\qbar},4_{\qbar})} \Big(B_4^0(1_q, 3_q, 4_{\bar q}, 2_{\bar q}) +  B_4^0(3_{ q}, 1_{q}, 2_{\bar q}, 4_{\bar q}) \\
&\quad + \hat{B}_4^0(1_q,3_q,4_{\bar{q}},2_{\bar{q}})
    - \frac{2}{N_C} \left[C_4^0(1_q,3_q,4_{\bar q},2_{\bar q}) + C_4^0(2_{\bar q},4_{\bar q},3_q,1_q)\right] \Big)
\end{split}
\label{eq:kIqq}
\end{equation}
where $C_4^0$ is the same-flavour four-quark interference function given in ref.~\cite{Gehrmann-DeRidder:2005btv}. (For a strict leading-colour treatment one could leave out the $C_4^0$ terms, but in the context of reducing matching uncertainties as much as possible we may as well include them here.) Following the notation of ref.~\cite{Gehrmann-DeRidder:2005btv}, the arguments in $B_4^0$ make the singularity explicit as $3_q || 4_{\bar q}$ in the first term and $1_q || 2_{\bar q}$ in the second.

Additionally, for each Born flavour channel, the matched NLO 3-jet rate must incorporate the $k_I$ correction factor as well, inclusive over all flavours. 
Thus, we apply a different $\calV_3^{\mathrm O}$ depending on the generated Born flavour channel. That flavour-dependent $\calV_{3q}^{\mathrm O}$ is given by:

\begin{eqnarray}\label{eq:V3q}
\calV_{3q}^{\mathrm O} & = & 2 \mathrm{Re}[M_{3q}^1 M_{3q}^{0*}] \,+ \,\int_{\hat t_3 > t_4} \d \Phi_4 \Bigg[ \calB_2^q \,(2 g^4 C_F) \, N_C \,A_{4}^0 + \calB_4^{q\textrm{(id.)}} \\[1mm] & & \hspace*{5.5cm} +~ \sum_{q'\neq q}^{n_f} k_{I}^{qq'}  \calB_4^q \Bigg]\,\delta^{(5)}\Big({\Phi}_3 - \mapPhi_3(\Phi_4)\Big)\,.\nonumber
\end{eqnarray}

We follow up with three additional comments on this procedure.

\begin{enumerate}
\item We note that, since all the terms in the denominator of $k_{I}^{qq'}$ are positive definite, there is no risk of accidental cancellations causing vanishing denominators here, and since the numerator is also positive definite, the factor $k_I$ is strictly positive and bounded.
\item In the following we focus on the $N_F$ piece of $\calV_{3q}^{\textrm{O}}$ only. Both the identical-flavour $1/N_C$ correction in eq.~\eqref{eq:kIqq} as well as the mixed-coupling interference term in eq.~\eqref{eq:Born-kI} yield finite contributions to $\calV_{3q}^{\mathrm{O}}$. This is because the $3$-parton configuration is resolved by construction, and both contributions yield finite results. This can also be seen by inspecting the structure of virtual corrections, see appendix~\ref{app:virtuals}, and realising that in the quark sector the singularity is solely proportional to $N_F$.
\item The definition of ``unclusterable" Born partons is valid not only for quark flavours, but also for Born gluons. Thus, in a $H\to gg$ Born-level event, no clusterings to states involving fewer than two gluons would be considered by the sector algorithm. Or, if one were to start from $Z\to q\bar{q}g$ events for some reason, no clusterings to states involving fewer than one gluon and one quark-antiquark pair of the requisite flavour would be considered. This makes the (unmatched) shower off a given Born state \emph{independent} of the existence of any other Born states, i.e., universal, which we consider a desirable property.\footnote{In the context of matching across different Born states, it can also be instructive to consider matching across $H\to b\bar{b}$, $c\bar{c}$, and in principle also $s\bar{s}$ and lighter-quark Born states since, unlike the $Z$ case, these have greatly different Born-level couplings. One would like a procedure that remains well-behaved and stable also for such cases, with matching factors mostly of order unity. (In the case of Higgs decays, $H\to gg$ Born states would of course also need to be included for a complete matching, and contributions from hadronic $H\to VV^*$ channels could be relevant too. Here, we restrict our attention to decays involving only $q\bar{q}H$ couplings.)} The matching then accounts for interferences between them, up to the matched order.
    
\end{enumerate}

\subsubsection{Sector resolution for $g\to q\bar{q}$ splittings}

As was shown in the first sector-shower study~\cite{Lopez-Villarejo:2011pwr}, it is necessary to pick a slightly modified form of the sector resolution variable for $g\to q\bar{q}$ splittings. The reason is that, while sector antenna functions for gluon emissions are symmetrised so that they contain also the collinear terms of the neighbouring antenna (and hence the full, symmetric $p_{gg}(x)$), the equivalent for a $gg\to \bar{q}qg$ antenna would have to include the collinear terms from a neighbouring $qg$ gluon-emission antenna.
Instead, the choice made in sector-shower implementations so far has therefore been to keep only the genuine $g\to q\bar{q}$ gluon-splitting terms in the gluon-splitting antennae. But one must then modify the sector boundaries so that the neighbouring $qg$ sector covers the entire $qg$-collinear region. 

This is achieved by defining the sector resolution variable as follows~\cite{Lopez-Villarejo:2011pwr,Brooks:2020upa}:
\begin{equation}
Q_\mathrm{res}^2 = \left\{
\begin{array}{cl}
\frac{s_{ij}s_{jk}}{s_{IK}} = t_{[j]/ik}\,, & \quad\mbox{if $j$ is a gluon}\\[2mm]
s_{ij}\sqrt{\frac{s_{jk}}{s_{IK}}}\,, & \quad\mbox{if $i,j$ is a same-flavour $q\bar{q}$ pair}
\end{array}
\right.\,,
\end{equation}
where $s_{ij}\equiv 2p_i\cdot p_j$.

\subsubsection{Sector subtraction for $g\to q\bar q$}
In this section we present the matching formula in the sector subtraction construction following the same steps as in \secref{sec:sectorsub}. In the $N_F$ channel we have the following
\begin{equation}\label{eq:IAsectornf}
I^{A,N_F}_{3\mapsfrom 4} = \calB_3
\frac{g^2 N_F}{A_3^0(\hat{1}_q,\hat{3}_g,\hat{2}_{\Bar{q}})}\int_{\hat{t}_3 > t_{4A}} \d\Phi_{+1}^A \,\Theta\big( Q^2_{{\rm res}, B} - Q^2_{{\rm res}, A}\big) \,B_4^0 \, .
\end{equation}
In the case of quarks, as explained above, the sector resolution variable $Q^2_{\rm res}$ is different than that of gluon emissions. Nevertheless, at the 4-quark level, the \textit{sector constraint} $\Theta\big( Q^2_{{\rm res}, B} - Q^2_{{\rm res}, A}\big)$ is identical to the gluon emission case $\Theta(t_{4B} > t_{4A})$ because we consider only massless quarks. Therefore we use the same notation $t_{4A/B}$.

\begin{figure}[h]
\centering
\includegraphics*[width=0.49\textwidth,page=2]{figures/plot-Zqq-2d-NF.pdf}
\includegraphics*[width=0.49\textwidth,page=1]{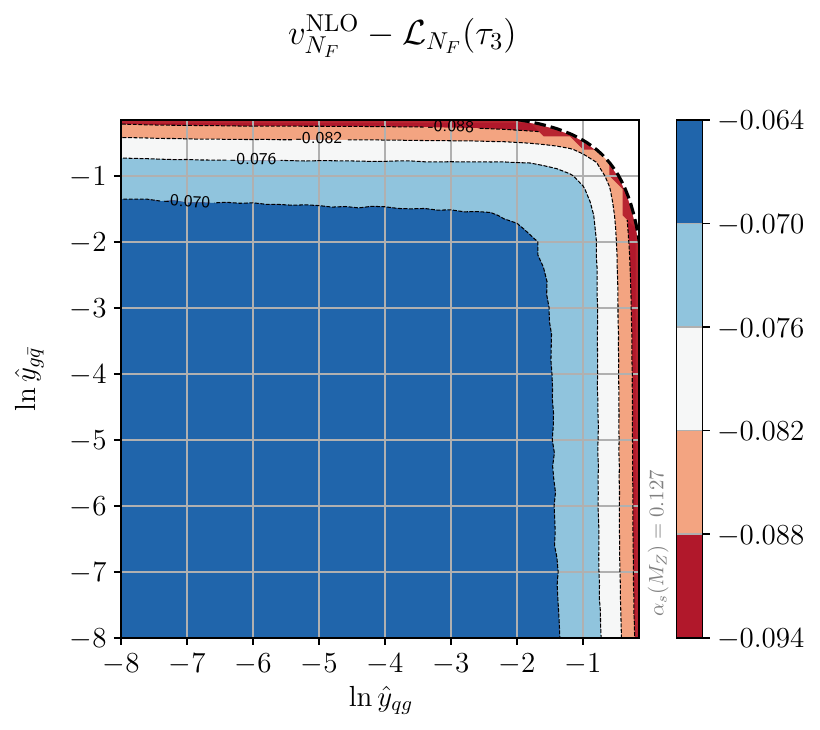}
\caption{The result of the $\KNLO_{N_F}$ factor as defined in eq.~\eqref{eq:KNLO-NF}, where we subtract the log-enhanced terms of eq.~\eqref{eq:wNF-expand}, $\mathcal{L}_{N_F}(\tau_3) = \frac{\alpha_s N_F}{2\pi}  \frac13 \ln\tau_3$.}

\label{fig:KNLO-nf}
\end{figure}

In order to construct a proper counter-term, we just need to realise that in the iterated (ordered) $2\mapsto 3$ limit, we have a $q\bar{q} \to q\bar{q}g$ splitting followed by a $g \to q^\prime \bar{q}^\prime$ splitting. Therefore, a proper counter-term must recover the full splitting function $p_{qg}$, i.e.
\begin{align}
    p_{qg}(z,\epsilon) = \frac{z^2 + (1-z)^2 - \epsilon}{(1-\epsilon)} \, .
\end{align}
We note that $p_{qg}$ contains {\em twice} the pole structure found in the real-virtual term. Similarly to the case of gluon emission, we restrict the subtraction term to the singular region $y_{13}<1/2$, which is sufficient to isolate the correct pole structure. We construct the following counter-term, which encodes the full $p_{qg}(z,\epsilon)$ splitting function,\footnote{Our counter-term differs from the $3$-parton tree-level antenna $E_3^0$ of refs.~\cite{Gehrmann-DeRidder:2005svg,Gehrmann-DeRidder:2005btv} only by finite terms.}
\begin{align}\label{eq:E30def}
    E_3^{s0}(1_q,3_{q^\prime},4_{\bar{q}^\prime}) = \frac{1}{s_{134}} \left( \frac{y_{13}^2 + y_{14}^2 - \epsilon}{(1-\epsilon) y_{34}} + y_{13} + y_{14} \right) \, .
\end{align}
We can now write down the equivalent of eq.~\eqref{eq:RR-sub-final-A}, {\em viz.}
\begin{align}\label{eq:RR-sub-final-A-NF}
\nonumber
I_{3\mapsfrom 4}^{A,N_F} &= \calB_3
 \left(\frac{\alpha_s N_F}{2\pi}\right)\, \left(\frac{\hat{y}_{qg}^{-\epsilon} }{2} \left( - \frac{1}{3\epsilon} - \frac{13}{12} + \frac{\ln 2}{3}  \right) - \mathcal{F}^{N_F}(\hat{y}_{g\Bar{q}}) \right) \\
&+\calB_3 \left(\frac{\alpha_s N_F}{2\pi}\right)\, \left( w^{A,N_F}_{\rm RS}(\hat{y}_{qg},\hat{y}_{g\Bar{q}})+ \mathcal{I}^{A,N_F}(\hat{y}_{qg},\hat{y}_{g\Bar{q}}) \right) \, ,
\end{align}
where we defined:
\begin{align}\label{eq:wRS-NF}
    w^{A,N_F}_{\rm RS} =\frac{8\pi^2}{A_3^0(\hat{1}_q,\hat{3}_g,\hat{2}_{\Bar{q}})}\int_{\hat{t}_3>t_{4A}} \d\Phi_{+1}^A \Theta\big( t_{4B} - t_{4A}\big) \bigg[ B_4^0 - \, A_3^0(\hat{1}_q,\hat{3}_g,\hat{2}_{\Bar{q}})\, E_3^{s0} \,\Theta\bigg(y_{13} < \frac12\bigg) \bigg] \, ,
\end{align}
and 
\begin{equation}\label{eq:Ifunction-NF}
\mathcal{I}^{A,N_F}
= 8 \pi^2 \int_{\hat{t}_3>t_{4A}}\d\Phi_{+1}^A \, E_3^{s0} \,\Theta\bigg(y_{13} < \frac12\bigg)  \bigg[\Theta\big( t_{4B} - t_{4A}\big) - 1\bigg] \, .
\end{equation}
Therefore, in complete analogy to eq.~\eqref{eq:KNLO}, we write
\begin{align}\label{eq:KNLO-NF}
\KNLO_{N_F} = \left(\frac{\alpha_s N_F}{2\pi}\right) \left( w^{A,N_F}_{\rm RS}+ \mathcal{I}^{A,N_F} + w^{A,N_F}_{\rm RS}+ \mathcal{I}^{A,N_F} + w_{\mathcal{V}}^{N_F}\right) \, ,
\end{align}
with the results plotted in fig.~\ref{fig:KNLO-nf}. The logarithmic structure of $w_{\mathcal{V}}^{N_F}$ has the following form
\begin{align}\label{eq:wNF-expand}
    \lim_{\tau_3 \to 0} w^{N_F}_{\mathcal{V}} \to \frac13 \ln\tau_3 + \frac{\ln 2}{3}  - \frac{11}{72}\, .
\end{align}

\subsection{The renormalisation terms}\label{sec:muR}

We are now ready to present the final NLO coefficient $A_{2\mapsto 3}^{\rm NLO}$, which enters our matching formula. Substituting eqs.~\eqref{eq:V2} and~\eqref{eq:V3Ofinal} in eq.~\eqref{eq:A3NLOb}, we have
\begin{align}
    A_{2\mapsto 3}^{\rm NLO} &= A_{2\mapsto 3}^{\rm LO} \, K^{\rm NLO} \, , \\
    K^{\rm NLO} &= \frac{1 - \Delta_2^1(\mZ^2,\hat{t}_3) + v^{{\rm NLO}}_{N_C} + v^{{\rm NLO}}_{N_F}}{1 + \frac{\alpha_s(\mZ)}{\pi}} \, ,
\end{align}
where the denominator comes from our choice to exactly reproduce the $\calB_2 + \calV_2$ denominator in eq.~\eqref{eq:A3NLOa},\footnote{This choice differs from that of ref.~\cite{Campbell:2021svd} at order $\mathcal{O}(\alpha_s^2)$ in $K^{\rm NLO}$, i.e.~beyond the matched accuracy.} and we emphasise that $\Delta_2^1$ in the numerator should be expanded to leading colour since $v_{N_C}^\mathrm{NLO}$ likewise does not include $1/N_C^2$ corrections.   

In addition to the above, we need to account for the scale and scheme choices used for the strong coupling on the shower side. The iterated $2\mapsto 3$ shower evaluates the coupling at the scale proportional to the transverse momentum of the emission, $\hat t_3 = \hat{y}_{qg} \hat{y}_{g\bar{q}} \mZ^2 = \tau_3 \mZ^2$, and additionally typically uses the CMW scheme~\cite{Catani:1990rr}:
\begin{equation}
    \alpha_s \to \alpha_s(\hat t_3) \left(1 + \frac{\alpha_s(\hat t_3)}{2\pi} K_{\textrm{CMW}}\right)\,.
\end{equation}
On the fixed-order side, it is customary to set the renormalisation scale $\mu_M = \mZ$.
We therefore need to subtract these two contributions in the NLO matching coefficient,
\begin{align}
    K^{\rm NLO}_{\textrm{full}} &= \frac{1 - \Delta_2^1(\mZ^2,\hat{t}_3) + \KNLO_{N_C} + \KNLO_{N_F} - \frac{\alpha_s}{2\pi} K_{\rm CMW} + \frac{\alpha_s}{2\pi} b_0 \ln\tau_3}{1 + \frac{\alpha_s(\mZ)}{\pi}} \, ,
    \label{eq:KNLO-full}
\end{align}
with the two-loop cusp anomalous dimension and the one-loop coefficient of the $\beta$-function,
\begin{align}
  K_{\rm CMW} = N_C \left(\frac{67}{18} - \frac{\pi^2}{6} \right) - \frac{5}{9} N_F, \quad b_0 = \frac{11 N_C}{6} - \frac{N_F}{3}\, .  
\end{align}
Inspection of eqs.~\eqref{eq:Sudontop-expand}, \eqref{eq:wNC-expand} and \eqref{eq:wNF-expand} reveals that in the limit $\tau_3 \to 0$ all $\ln\tau_3$ enhanced contributions cancel out in the result for $K^{\rm NLO}_{\textrm{full}}$.

We note that the scale and scheme choice used for $\alpha_s$ in the numerator of $K^\mathrm{NLO}_{\mathrm{full}}$ is an ${\cal O}(\alpha_s^3)$ ambiguity and is hence not fixed by the NNLO matching conditions. Either of the choices $\mu^2 = \mu_M^2 = \mZ^2$ or $\mu^2 = \hat t_3$ could be justified. As a pragmatic solution, pending further investigations, we advocate using the geometric mean, $\mu^2 = \sqrt{\hat t_3} \mZ$, with the two 
limiting values used for uncertainty estimates. 

A further subtlety, which was also touched on in ref.~\cite{Hartgring:2013jma}, is what happens as the perturbative evolution in the shower crosses flavour thresholds, i.e., for $\hat t_3 < m_b^2$ and for $\hat t_3 < m_c^2$. In the VINCIA shower, at each flavour threshold,
\begin{enumerate}
\item The phase space for gluon splittings, $g\to Q\bar{Q}$, closes. Hence the value of  $N_F$ for real splittings decreases by one. 
\item The value of $N_F$ used in the running of $\alpha_s$ (including in the evaluation of the CMW factor) decreases by one. The running is matched across flavour thresholds so that it is continuous. Thus, $\alpha_s^{N_F=5}(m_b^2) = \alpha_s^{N_F=4}(m_b^2)$ and $\alpha_s^{N_F=4}(m_c^2) = \alpha_s^{N_F=3}(m_c^2)$.
\end{enumerate}

Consider the $N_F$-dependent term in the combination of the one-loop and real matrix elements in $\calV_3^{\mathrm{O}}$, 
\begin{equation}
\frac{\alpha_s(\mZ^2)}{2\pi}\frac{N_F}{3}\ln\left[ \frac{\hat t_3}{ \mZ^2}\right] \calB_3\,,
\end{equation}
with $N_F = 5$ for a renormalisation scale $\mu_M = \mZ$. For values of the transverse momentum $\sqrt{\hat t_3}$ below the $b$-quark mass, we ``freeze'' the contribution from the $b$ quark, 
\begin{equation}
{\hat t_3 < m_b^2~:} ~~~\frac{\alpha_s(\mZ^2)}{2\pi}\left(\frac{4}{3}\ln\left[ \frac{\hat t_3}{ \mZ^2}\right] + \frac13\ln\left[ \frac{m_b^2}{\mZ^2}\right]\right)\calB_3\,,
\end{equation}
and similarly, below the $c$-quark threshold,
\begin{equation}
{\hat t_3 < m_c^2~:} ~~~\frac{\alpha_s(\mZ^2)}{2\pi}\left(\frac{3}{3}\ln\left[ \frac{\hat t_3}{ \mZ^2}\right] 
+ \frac13\ln\left[ \frac{m_b^2}{\mZ^2}\right]
+ \frac13\ln\left[ \frac{m_c^2}{\mZ^2}\right]\right)\calB_3\,.
\end{equation}
This is equivalent to the structure that results from changing the running of the coupling across flavour thresholds, as is done by default in VINCIA. There is therefore no remainder from this modification.  

\section{Uncertainties}
\label{sec:uncertainties}

Any calculation is only as good as its uncertainty estimates. In the matching between NNLO perturbation theory and the sector shower, there are several sources of ambiguities which generate uncertainties starting from ${\cal O}(\alpha_s^3)$ and ${\cal O}(\alpha_s^2/N_C^2)$. (The latter is due to the fact that we did not here include the virtual corrections at full colour.)

Starting with NLO matching, there are the well-known ambiguities between MC@NLO-type~\cite{Frixione:2002ik} and \powheg-type~\cite{Frixione:2007vw} matching schemes. In our notation, an analogous difference can be induced by introducing a parameter ${\rho}\in[-1,2]$ in the definition of the tree-level matched antenna function,
\begin{equation}
A_{2\mapsto 3}^\mathrm{LO} ~~\to~~ A_{2\mapsto 3}^\mathrm{LO}({\rho}) ~=~A_{2\mapsto 3}^\mathrm{LO}\,\frac{\calB_2}{\calB_2 + {{\rho}} \calV_2}\,,
\end{equation}
where $\rho=0$ yields the functional form of the matching from ref.~\cite{Campbell:2021svd}.
Generalising this to NNLO matching the ambiguity moves one order higher, 
\begin{equation}
A_{2\mapsto 3}^\mathrm{NLO}~~\to~~
A_{2\mapsto 3}^\mathrm{NLO}(\rho)~=~A_{2\mapsto 3}^\mathrm{NLO} \,\frac{\calB_2 + \calV_2}{\calB_2 + \calV_2 + \rho_{\calW} \calW_2}\,.
\end{equation}

We also introduce explicit ${\cal O}(\alpha_s^3)$ ambiguities in the definitions of $\calV_3$ and the tree-level four-parton antenna functions, via substitutions
\begin{eqnarray}
\calV_3 & \to & 
\calV_3 \,\pm\, \left(\frac{\alpha_s}{2\pi}\right)^3\!\rho_{\calV} \,, \\[1mm]
\calB_4 & \to & 
\calB_4 \,\pm\, \left(\frac{\alpha_s}{2\pi}\right)^3\! \rho_{\calT}\,,
\end{eqnarray}
with $\rho_{\calV,\calT} = {\cal O}(1)$. 

For the scale definitions in the $\alpha_s$ evaluations here, we again recommend taking the geometric mean of the pre- and post-branching scales as the central value, i.e., $\mu^2 =\sqrt{t_3} \mZ$ for $\calV_3$ and $\mu^2 = \sqrt{t_3 t_4}$ for $\calB_4$ (and $\mu^2=t_4$ for direct $2\mapsto4$ phase-space points). In addition to this, the renormalisation scale used for the explicit $\alpha_s$ factors in $K^\mathrm{NLO}$ can be varied between $t_3$ and $\mZ$, as was discussed above. 

Finally, we propose to match on to conventional renormalisation-scale variations in the pure shower description by varying the renormalisation scale used to evaluate $\alpha_s$ in $A_{3\mapsto4}$ and $A_{2\mapsto 4}$ by a factor $\propto 2$ around $t_4$. 

\section{Summary \& Outlook}
\label{sec:conclusions}

In this paper we derived the necessary matching conditions to achieve NNLO accuracy within the VINCIA sector shower algorithm, with the NLO 3-jet correction expanded to leading colour. We have built on the consistent inclusion of direct $2\mapsto 4$ branchings in the unordered phase space of the sector shower, following ref.~\cite{Li:2016yez}. In particular, we utilised the sector resolution variable to cleanly separate out the ordered phase space (populated by iterated $2\mapsto 3$ branchings) from the unordered one (produced by direct $2\mapsto 4$ branchings). The fixed-order jet rates used in the matching conditions were defined according to the ARCLUS $3 \mapsto 2$ clustering algorithm~\cite{Lonnblad:1992qd}. This is especially convenient because VINCIA's sector resolution criterion aligns with the ARCLUS jet resolution.

The main result of this paper is the derivation and computation of the $K^{\textrm{NLO}}$ 3-jet matching coefficient given in eq.~\eqref{eq:KNLO-full}. The major difficulty in the evaluation of this factor consists in the presence of the sector constraint in the 4-parton phase space. We devised a subtraction procedure optimised to achieve high accuracy on the numerical component of $K^{\textrm{NLO}}$. This enables the matching coefficient to be efficiently included in VINCIA, thereby delivering the last missing ingredient for NNLO matching in our shower.

In a follow-up paper, we plan to present phenomenological results for hadronic $Z$ decays, in combination with a public release of the VINCIA code implementing our method.
We believe that our procedure can be generalised to more elaborate processes but this will require additional work, e.g., to extend the matching with subleading-colour effects, initial-state radiation, and final states with more than two Born legs.

\section*{Acknowledgements}

We thank J.~Altmann for comments on the manuscript. BKE and PS are supported by the Australian Research Council under grant DP220103512. LS is supported by the Australian Research Council under grant DE230100867. 
We are grateful to the Rudolf Peierls Centre for Theoretical Physics, University of Oxford, and to the Aspen Center for Physics, Colorado, for hospitality and stimulating working environments during discussions of this work. In Oxford, PS was supported by a Royal Society Wolfson Visiting Fellowship RSWVF$\backslash$R1$\backslash$231006. 
The Aspen Center for Physics is supported by National Science Foundation grant PHY-2210452.

\appendix 


\section{Dissecting sector constraints in VINCIA}\label{app:secanal}

We shall demonstrate the important features of the sector constraint using a generic antenna map, e.g.~for a $qg$ antenna,
\begin{align}
    p_1 &= a_1 \,\hat{p}_q + b_1 \,\hat{p}_g + c\, p_\perp \, , \\
    p_3 &= a_3 \,\hat{p}_q + b_3 \,\hat{p}_g - p_\perp \, , \\
    p_4 &= a_4 \,\hat{p}_q + b_4 \,\hat{p}_g +(1-c)\, p_\perp \, ,
\end{align}
where the variable $p_\perp$ is the standard transverse momentum in the Sudakov parametrisation (it is spacelike and orthogonal to $\hat{p}_q$ and $\hat{p}_g$). We note that it is simply a parametric variable which should not be confused with the shower evolution variable, which we denote by $t$. 
The quantities $a_i$, $b_i$ and $c$ encode the explicit kinematic map and are a function of $(y_{13},y_{34})$. We emphasise that this representation is general for local antenna kinematics; in particular all of the maps so far implemented in VINCIA (i.e., the so-called ARIADNE~\cite{Lonnblad:1992tz}, Kosower~\cite{Kosower:2003bh}, and dipole~\cite{Catani:1996jh} maps) correspond to different functional forms of $c(y_{13},y_{34})$.
Nevertheless, the latter has to satisfy the following conditions in the quark-gluon and gluon-gluon collinear limits,
\begin{align}\label{eq:climits}
    \lim_{y_{13} \to 0} c(y_{13},y_{34}) = 1, \quad \lim_{y_{34} \to 0} c(y_{13},y_{34}) = 0 \, .
\end{align}
From energy-momentum conservation and on-shell conditions we also have
\begin{align}
    a_1 + a_3 + a_4 = b_1 + b_3 + b_4 = 1 \, , \quad  -p^2_\perp = p_t^2 = a_3 b_3 \hat{s}_{qg}  \, .
\end{align}
It is clear that the sector constraint in eq.~\eqref{eq:IAsector} explicitly depends on the anti-quark momentum $p_2$. For local antenna kinematics, $p_2$ does not participate in the $4 \mapsto \hat{3}$ clustering for sector $A$, i.e., $p_2 = \hat{p}_{\bar{q}}$. To proceed, we introduce a Sudakov decomposition of $p_2$ in terms of the Born$+1$ momenta $(\hat{p}_q,\hat{p}_g)$:
\begin{align}
    p_2 = a_2 \,\hat{p}_q + b_2 \,\hat{p}_g + p_{\perp 2} \, ,
\end{align}
and it is straightforward (by on-shellness) to derive expressions for the Sudakov coefficients in terms of Born$+1$ invariants
\begin{align}
    a_2 = \frac{\hat{y}_{q\Bar{q}}}{\hat{y}_{qg}}, \quad b_2 = \frac{\hat{y}_{g\Bar{q}}}{\hat{y}_{qg}} , \quad p_{t2}^2 = a_2 b_2 \hat{s}_{qg} \, ,
\end{align}
with $\hat{y}_{ij} = \hat{s}_{ij}/\mZ^2$. 

From this last step, we can glean the important features of the sector constraint, and we do so by inspecting its azimuthal dependence. The azimuthal angle, $\varphi$, in the antenna phase space, eq.~\eqref{eq:antPS4D}, can be chosen to be the relative angle between the 2-dimensional transverse vectors $p_\perp$ and $p_{\perp 2}$. Therefore, in terms of the antenna phase-space variables the sector constraint can be rearranged as follows
\begin{align}
    \Theta\big( t_{4B} - t_{4A}\big) = \Theta\big( f(y_{13},y_{34};\hat{y}_{qg},\hat{y}_{g\Bar{q}},\hat{y}_{q\Bar{q}}) - \cos\varphi\big) \, ,
\end{align}
where the function $f$ depends explicitly on the specific map, as defined through the choice of $c(y_{13},y_{34})$. We obtain
\begin{align}\label{eq:f}
f(y_{13},y_{34};\hat{y}_{qg},\hat{y}_{g\Bar{q}},\hat{y}_{q\Bar{q}}) = \frac{(1-y_{13}) (a_2 b_4 + a_4 b_2) - y_{13} (a_2 b_3 + b_2 a_3) - y_{13} y_{34}}{2\sqrt{\frac{\hat{y}_{q\Bar{q}} \hat{y}_{g\Bar{q}}}{\hat{y}_{qg}^2}} (1-(1-y_{13})c) \sqrt{a_3 b_3}} \,.
\end{align}

Thus the sector constraint boils down to two distinct regions in $\varphi$: one in which there is no constraint on $\varphi$, corresponding to $f > 1$, and one in which the $\varphi$ range is bounded, in the region where $|f|<1$:
\begin{align}\label{eq:secconsphi}
   \Theta\big( t_{4B} - t_{4A}\big) = \Theta\big( f > 1 \big) \Theta\left( 0 < \varphi < \pi\right) ~+~ \Theta\left(-1 < f <1 \right) \,\Theta(\varphi < \cos^{-1}(f) )
 \, .
\end{align}
%

To study the behaviour of $f$ in the singular limits, we utilise the explicit expressions of the Sudakov coefficients from ref.~\cite{Preuss:2024vyu} and eq.~\eqref{eq:climits}.
In the limit $y_{13} \to 0$, we obtain 
\begin{align}
    \lim_{y_{13} \to 0} f(y_{13},y_{34};\hat{y}_{qg},\hat{y}_{g\Bar{q}},\hat{y}_{q\Bar{q}}) = + \infty \,,
\end{align}
and so, as $y_{13} \to 0$, the $\varphi$-integral is unconstrained as per eq.~\eqref{eq:secconsphi}. The limit $y_{34} \to 0$ is more interesting. At fixed $y_{13}$, we have two cases:
\begin{equation}
    \lim_{y_{34} \to 0} f(y_{13},y_{34};\hat{y}_{qg},\hat{y}_{g\Bar{q}},\hat{y}_{q\Bar{q}}) =
    \begin{cases}
    +\infty\,, & \text{if}\  y_{13}<\frac12 \\
    -\infty\,, & \text{if}\  y_{13}>\frac12 \,.
    \end{cases}
\end{equation}

Therefore, from eq.~\eqref{eq:secconsphi} we conclude that the region $y_{13}>\frac12$ is completely regular (i.e., free of singularities) because the sector step function in this limit vanishes. In other words, the sector constraint shields the singularity $y_{34}\to 0$ in the region $y_{13}>1/2$ of the Dalitz space. 

\section{The function $\mathcal{F}(x)$}
\label{app:Ffunction}
Here we list the finite function that appears in eq.~\eqref{eq:D30wosector}, which results from the ordering restriction $t_{4A} < \hat{t}_3$. We first define an auxiliary function
\begin{align}
\nonumber
 \mathcal{G}^{N_C}(y) = \frac{31 y}{6} + \frac{3 y^2}{4} - \frac{y^3}{9} + \frac{y^4}{16} + \frac{\pi^2}{3} - \frac{11 \tanh^{-1}(y)}{3} - 2\ln^2 2 + \frac32 \ln(1-y) \\
 +\ln^2(1-y) - 2\ln(1-y)\ln(1+y)  + \ln\left(\frac{16}{1+y}\right) \ln(1+y) - 4\,{\rm Li}_2\left(\frac{1+y}{2}\right) \, ,
\end{align}
and thus
\begin{align}
    \mathcal{F}^{N_C}(x) = \frac12 \Theta\left(x \leq \frac14\right) \mathcal{G}^{N_C}(\sqrt{1-4x}) \, .
\end{align}
Similarly for the $N_F$ contribution we obtain
\begin{align}
\nonumber
 \mathcal{G}^{N_F}(y) = \frac{1}{72}\bigg( y(- 66 - 9 y + 8 y^2) + 6(3y^2-7)\ln(1-y)+24\ln(1+y)\bigg) \, .
\end{align}
\section{Virtual corrections}\label{app:virtuals}
Using the notation of ref.~\cite{Gehrmann-DeRidder:2005btv}, virtual corrections can easily be obtained from the 1-loop $qg\bar{q}$ antenna function, $A_3^1$, and the integrated one-loop $q\bar{q}$ antenna function, $\mathcal{A}_2^1$. We can express the real-virtual contribution as follows:
\begin{multline}
\nonumber
    2 \mathrm{Re}[M_3^1 M_3^{0*}] = \calB_3 \left(\frac{\mu^2_R}{\mZ^2}\right)^{\epsilon} \mZ^{2\epsilon} \, \bigg[\left(\frac{\alpha_s N_C}{2\pi}\right)\left(\frac{A_3^1(\hat{1}_q,\hat{3}_g,\hat{2}_{\Bar{q}})}{A_3^0(\hat{1}_q,\hat{3}_g,\hat{2}_{\Bar{q}})} + \mathcal{A}_2^1\left(\mZ^2\right)  \right) \\
    + \left(\frac{\alpha_s N_F}{2\pi}\right)\frac{\hat{A}_3^1(\hat{1}_q,\hat{3}_g,\hat{2}_{\Bar{q}})}{A_3^0(\hat{1}_q,\hat{3}_g,\hat{2}_{\Bar{q}})}\bigg] \, ,
\end{multline}
where $A_3^1$ is the $q\bar{q} g$ antenna function at one loop, $\mathcal{A}_2^1$ is the integrated one-loop $q\bar{q}$ antenna function and $\hat{A}_3^1$ is the fermionic one-loop contribution.
It is best to define the following dimensionless functions,
\begin{multline}
    T_3^0(\yqg,\yqbg) = \frac{2(1-\yqg-\yqbg)}{\yqg \yqbg} + \frac{\yqg}{\yqbg} + \frac{\yqbg}{\yqg} \, , \\
    T_3^1(\yqg,\yqbg) = \frac{1}{2 \yqg} - \frac{\yqg}{2(1-\yqbg)}
    + \ln(\yqg)\left(2-\frac{\yqg \yqbg}{2(1 - \yqg)^2} + 2 \frac{\yqg- \yqbg}{1-\yqg}\right) + \yqg \leftrightarrow \yqbg \, .
\end{multline}
Using the relevant expressions from \cite{Gehrmann-DeRidder:2005btv} we obtain
\begin{multline}
    2 \mathrm{Re}[M_3^1 M_3^{0*}] =\\
     \calB_3 \left(\frac{\alpha_s N_C}{2\pi}\right) \left(\frac{\mu^2_R}{\mZ^2}\right)^{\epsilon} \bigg[-\frac{2}{\epsilon^2} - \frac{10}{3\epsilon} + \frac{\ln \hat{y}_{qg} + \ln \hat{y}_{g\bar{q}}}{\epsilon} - 4 + \frac{7\pi^2}{6}
    - \frac12 \ln^2 \yqg - \frac12 \ln^2 \yqbg\\ 
    - R(\yqg,\yqbg) + \frac{T_3^1(\yqg,\yqbg)}{T_3^0(\yqg,\yqbg)}\bigg]
    + \calB_3 \left(\frac{\alpha_s N_F}{2\pi}\right) \left(\frac{\mu^2_R}{\mZ^2}\right)^{\epsilon} \frac{1}{3\epsilon}\, ,
\end{multline}
where $R(\yqg,\yqbg)$ is given in ref.~\cite{Gehrmann-DeRidder:2005btv}:
\begin{align}
    R(x,y) = \ln x \, \ln y - \ln x \ln(1-x) - \ln y \ln(1-y) + \frac{\pi^2}{6} - {\rm Li}_2(x) - {\rm Li}_2(y) \, .
\end{align}
\section{NLO 3-jet matching with global subtraction}
\label{app:globalsub}

As an alternative to the directly constructed sector subtraction presented in the main body of the paper, we also investigated a variant based on global antenna subtraction. In our studies, this turned out to be less numerically efficient, but we present it here for future reference. 

Using that the pole structure of $2\mathrm{Re}[M_3^1M_3^{0*}]$ should be reproduced by integrals over global quark-gluon antenna-subtraction terms, we first rewrite 
\begin{equation}
\int_{\hat{t}_3>t_{4A}} \hspace*{-4mm}\d\Phi_{+1}^A \Bigg(\Theta\big( t_{4B} - t_{4A}\big) \calB_4 ~-~ \calB_3 \, g^2 N_C \, d_3^A\Bigg)~+~{\calB_3\, g^2 N_C \overbrace{\int_{\hat{t}_3>t_{4A}} \hspace*{-4mm}\d\Phi_{+1}^A \,d_3^A }^{I^\mathrm{gO}_{3A}}}
~+~ A\leftrightarrow B~,\label{eq:I43sub}
\end{equation}
where $d^A_3$ can be chosen, e.g., to be the GGG sub-antenna function for gluon emission off a quark-gluon antenna~\cite{Gehrmann-DeRidder:2005btv} (modulo colour and coupling factors), {\em viz.}
\begin{align}
    d_3^A(1_q,3_g,4_g) = \frac{1}{s_{134}} \left(\frac{2y_{14}}{y_{13} y_{34}} + \frac{y_{34}}{y_{13}} - y_{34} + \frac{y_{14} y_{13}}{y_{34}} + \frac52 + \frac{y_{34}}{2} -\epsilon \frac{y_{34}}{y_{13}}\right) \, .
\end{align}
The subtraction term, i.e.~$I^\mathrm{gO}_{3A}$, then only differs from the standard global NLO antenna-subtraction integrals by the restriction to $\hat{t}_3 > t_{4A}$. It is given by:
\begin{eqnarray}
I_{3A}^\mathrm{gO} & = & g^2 N_C \int_{\hat{t}_3>t_4} \d\Phi^A_{+1} \,d_3^A \\[2mm]
& = & g^2 N_C \left[ \frac{{\cal D}^0_3(\epsilon,\hat{s}_{qg})}{2} - \int_{t_4>\hat{t}_3} \d\Phi^A_{+1} \,d_3^A\right] \\[2mm] \nonumber
& = & \frac{\alpha_s N_C}{2\pi}\left(-2{\bf I}_{qg}(\epsilon,\hat{s}_{qg}) \,+\,\frac{17}{3}\right) \label{eq:I3A}\\  \nonumber
&+&\,\frac{\alpha_s N_C}{2\pi}\Theta\left(\hat{y}_{g\bar{q}} <\frac14\right)\Bigg[  \frac{7x^3}{72} - \frac{285 x}{72} + \left( \frac{285}{72} - \frac{45}{72} x^2 \right) {\rm tanh}^{-1}(x)   \\   \nonumber
& & \hspace{24mm} + \frac12 \ln(1-x) \ln\frac{4}{1-x}+ \frac12 \ln\frac{1+x}{4} \ln(1+x) \\
& & \hspace{24mm}- {\rm Li}_2\left(\frac{1-x}{2}\right) + {\rm Li}_2\left(\frac{1+x}{2}\right)\Bigg]\,,
\end{eqnarray}
with $x = \sqrt{1-4\hat{y}_{g\bar{q}}}$.
The integrated tree-level antenna function, ${\cal D}_3^0$, and its decomposition into pole (${\bf I}_{qg}$) and finite terms is given in \cite{Gehrmann-DeRidder:2005btv}. 
The finite terms of eq.~\eqref{eq:I3A} that arise from the restriction $t_4 > \hat{t}_3$ would be absent in a standard fixed-order context.
The expression for the matching coefficient $\calV_3^\mathrm{O}$ is then
\begin{eqnarray}
\label{eq:V3Oglb}
\calV_3^\mathrm{O} & = &  {\color{black}\overbrace{\color{black}2\mathrm{Re}\big[M_3^1M_3^{0*}\big] + \bigg(I_{3A}^\mathrm{gO} + I_{3B}^\mathrm{gO}\bigg)\,\calB_3 }^{\mbox{finite}}} \\[2mm]
& & + \int_{\hat{t}_3>t_{4A}} \hspace*{-4mm}\d\Phi_{+1}^A \Bigg(\Theta\big( t_{4B} - t_{4A}\big) \calB_4 - \calB_3\,g^2 N_C\, d_3^A \Bigg)~+~A\leftrightarrow B \nonumber \, .
\end{eqnarray}

However, the terms on the second line in eq.~\eqref{eq:V3Oglb} cannot yet be integrated numerically. This is because the term featuring the global antenna function does not provide a {\em local} subtraction (point-by-point) in phase space. This is the main difference between this approach and that based on sector subtraction presented in the body of the paper.
We can still leverage the insights from \secref{sec:sectorsub} to rearrange the computation and make it amenable to numerical integration.
The analysis of the sector constraint in \secref{sec:sectorsub} proved that the gluon-gluon collinear singularity is contained in the region $y_{13}<1/2$ as depicted in fig.~\ref{fig:Dalitz}.

Given that the sub-antenna function $d_3^A$ represents a smooth partial-fractioning of $p_{gg}$, $d_3^A$ does not contain the full singular behaviour of $A_4^0$. Therefore, we can remove the gluon-gluon collinear limit of $d_3^A$ and replace it instead with the full splitting function $p_{gg}$ in the region $y_{13} < 1/2$. Therefore, we obtain:
\begin{eqnarray}
\nonumber
\calV_3^\mathrm{O} & = &  {\color{black}\overbrace{\color{black}2\mathrm{Re}\big[M_3^1M_3^{0*}\big] + \bigg(I_{3A}^\mathrm{gO} + I_{3B}^\mathrm{gO}\bigg)\,\calB_3 }^{\mbox{finite}}} \\[2mm]
&+& \calB_3\, g^2 N_C \int_{\hat{t}_3>t_{4A}} \hspace*{-4mm}\d\Phi_{+1}^A \left[\frac{1}{s_{34}} p_{gg}(y_{13}) \Theta\left(y_{13}<\frac12\right) - \tilde{d}_3^A \right] \nonumber \\[2mm] \nonumber
&+& \int_{\hat{t}_3>t_{4A}} \hspace*{-4mm}\d\Phi_{+1}^A \Bigg(\Theta\big( t_{4B} - t_{4A}\big) \calB_4 - \calB_3\,g^2 N_C \left[ \frac{1}{s_{34}} p_{gg}(y_{13})\,\Theta\left(y_{13}<\frac12\right) + (d_3^A - \tilde{d}_3^A) \right] \Bigg) \\ [2mm]
&+& A\leftrightarrow B \label{eq:V3Ofull} \, ,
\end{eqnarray}
where
\begin{align}
    p_{gg}(y_{13}) &= \frac{2y_{13}}{(1-y_{13})} + \frac{2(1-y_{13})}{y_{13}} + 2 y_{13} (1-y_{13}) \, ,\\ 
    \tilde{d}_3^A &= \frac{1}{s_{34}} \left(\frac{(1-y_{13})}{y_{13}} + y_{13} (1-y_{13}) \right) \, .
\end{align}
The last line in eq.~\eqref{eq:V3Ofull} is now amenable to numerical integration. The second line in eq.~\eqref{eq:V3Ofull} is finite, albeit it needs to be evaluated in $D = 4-2\epsilon$ to cancel the potential $y_{34} \to 0$ singularity before taking the limit $\epsilon =0$. Using the phase space given in eq.~\eqref{eq:PS-eps}, we obtain:
\begin{eqnarray}
\nonumber
\calV_3^\mathrm{O} & = &  {\color{black}\overbrace{\color{black}2\mathrm{Re}\big[M_3^1M_3^{0*}\big] + \bigg(I_{3A}^\mathrm{gO} + I_{3B}^\mathrm{gO}\bigg)\,\calB_3 }^{\mbox{finite}}} \\[2mm]
&+& \calB_3\, \frac{\alpha_s N_C}{2\pi}\, \frac12 \left( \frac{\pi^2}{6} - \frac{1}{24} - \frac{11 \ln2}{6}\right) \\ [2mm] \nonumber
&+& \int_{\hat{t}_3>t_{4A}} \hspace*{-4mm}\d\Phi_{+1}^A \Bigg(\Theta\big( t_{4B} - t_{4A}\big) \calB_4 - \calB_3\,g^2 N_C \left[ \frac{1}{s_{34}} p_{gg}(y_{13})\,\Theta\left(y_{13}<\frac12\right) + (d_3^A - \tilde{d}_3^A) \right] \Bigg) \\ [2mm]
&+& A\leftrightarrow B \nonumber \, .
\end{eqnarray}
Finally, we have checked explicitly that the above approach yields identical results to the sector subtraction approach presented in the body of the article.

\section{Implementation of the matching algorithm}
\label{app:matching-alg-descr}
In this section, we present the explicit algorithm used to implement our matching scheme. While a detailed validation is deferred to future work, our focus here is to highlight how the matched emission densities --- eqs.~\eqref{eq:A2to4}, \eqref{eq:A3to4LO}, and \eqref{eq:A3NLOb} --- are incorporated within the algorithm.

\renewcommand{\thealgorithm}{}
\begin{algorithm}[h!]
  \caption{Implementation of the matching scheme}
  \label{alg:sub-to-int}
  \begin{algorithmic}[1]
    \State Generate a Born-level phase space point with weight $\calB_2+\calV_2+\calW_2$
    \State Define the starting scale for the shower, $t_\mathrm{now} = m_Z^2$, as the squared invariant mass of the two Born partons
    \State\label{st:nextscale}Starting from the current shower scale, $t_\mathrm{now}$, let the $2\mapsto 3$ and $2\mapsto 4$ branchers compete for the highest branching scale 
    \State Update the current shower scale to be that of the winning branching, $t_\mathrm{now} = \mathrm{max}(t_{2\mapsto 3},t_{2\mapsto 4})$
    \If{$t_\mathrm{now} < t_c$}~exit the shower
    \EndIf
    \If{winning branching is a $2\mapsto 3$ branching}
       \State Calculate the acceptance probability including $A^\mathrm{NLO}_{2\mapsto3}$ in eq.~\eqref{eq:A3NLOa}
            \If{emission is rejected} go back to step \ref{st:nextscale}
            \Else
                \State\label{st:3to4} \parbox[t]{\dimexpr\linewidth-4em}{Generate a $3\mapsto4$ branching and include $A^\mathrm{LO}_{3\mapsto 4}$ when calculating accept probabilities for the $3\mapsto4$ step}
                    \vspace{0.1in}
                    \State Update the shower scale $t_{\rm now} = t_{3\mapsto 4}$
                    \If{$t_\mathrm{now} < t_c$}~exit the shower
                    \EndIf
                    \If{$3\mapsto 4$ branching is accepted} continue from step~\ref{st:rest-shower}
                    \Else~continue from step~\ref{st:3to4}
                    \EndIf
            \EndIf    
    \EndIf
    \If{winning branching is a $2\mapsto 4$ branching}
        \State Calculate the acceptance probability including $A^\mathrm{LO}_{2\mapsto4}$
        \If{rejected} continue from step~\ref{st:nextscale}
        \Else~continue from step~\ref{st:rest-shower}
        \EndIf   
    \EndIf
    \State\label{st:rest-shower}Continue with the standard shower from the resulting $4$-particle configuration
  \end{algorithmic}
\end{algorithm}

\bibliographystyle{JHEP}
\bibliography{bibliography}

\end{document}